\documentclass[10pt,journal,compsoc]{IEEEtran}
\ifCLASSINFOpdf
   \usepackage[pdftex]{graphicx}
\else
\fi
\usepackage{algorithmic}
\usepackage{url}

\usepackage{adjustbox}
\usepackage{amsmath,amssymb,amsfonts}
\usepackage{algorithmic}
\usepackage{textcomp}
\usepackage{xcolor}
\usepackage{comment}
\usepackage{color, colortbl}
\usepackage{caption}
\usepackage{subfig}
\usepackage{threeparttable}
\usepackage{pifont}

\usepackage[style=ieee, minnames=1, maxnames=1]{biblatex}
\definecolor{Gray}{gray}{0.9}

\newif{\ifSubmit}
\newif{\ifFinal}
\newif{\ifDraft}
\Drafttrue

\ifSubmit
\newcommand{\maic}[1]{}
\newcommand{\omcomment}[1]{}
\newcommand{\duocomment}[1]{}
\newcommand{\rzcomment}[1]{}
\newcommand{\tabacomment}[1]{}
\else
\newcommand{\maic}[1]{\noindent\textcolor{red}{MZ: #1}}
\newcommand{\omcomment}[1]{\noindent\textcolor{orange}{\bf Om: #1}}
\newcommand{\duocomment}[1]{\noindent\textcolor{blue}{\bf Duo: #1}}
\newcommand{\rzcomment}[1]{\textcolor{brown}{\textbf{RZ: #1}}}
\newcommand{\tabacomment}[1]{\textcolor{teal}{\textbf{Taba: #1}}}
\fi

\begin{document}
%

\title{PROV-IO$^+$: A Cross-Platform Provenance Framework for Scientific Data on HPC Systems }
%
%
%
%

\author{Runzhou Han$^\dagger$,
        Mai Zheng$^\dagger$,
        Suren Byna$^\ddagger$, 
        Houjun Tang$^\S$,
        Bin Dong$^\S$,
        Dong Dai$^\P$,  \\ 
        Yong Chen\textcolor{blue}{$^\ding{112}$}, 
        Dongkyun Kim\textcolor{green}{$^{\ding{112}}$}, 
        Joseph Hassoun\textcolor{green}{$^{\ding{112}}$},
        David Thorsley\textcolor{green}{$^{\ding{112}}$},
        Matthew Wolf\textcolor{blue}{$^\ding{112}$} \\
        \normalsize $^\dagger$\textit{Department of Electrical and Computer Engineering, Iowa State University}\\
        $^\ddagger$\textit{Department of Computer Science and Engineering, The Ohio State University}\\
        $^\S$\textit{Scientific Data Management Group, Lawrence Berkeley National Laboratory}\\
        $^\P$\textit{Computer Science Department, University of North Carolina Charlotte}\\
        \textcolor{blue}{$^\ding{112}$} \textit{Systems Architecture Lab, Samsung} \hspace{8pt}
        \textcolor{green}{$^{\ding{112}}$} \textit{Samsung Advanced Institute of Technology, Samsung}
        }

\IEEEtitleabstractindextext{%
\begin{abstract}
Data provenance, or data lineage, describes the life cycle of data. In scientific workflows on HPC systems, scientists often seek diverse provenance (e.g., origins of data products,  usage patterns of datasets). 
Unfortunately, existing provenance solutions cannot address the challenges due to their incompatible provenance models and/or system implementations.  

In this paper, we analyze four representative scientific workflows in collaboration with the domain scientists to identify concrete provenance needs. 
Based on the first-hand analysis, we propose a provenance framework called PROV-IO$^+$, which includes an I/O-centric provenance model for describing scientific data and the associated I/O operations and environments precisely.  
Moreover, we build a prototype of PROV-IO$^+$ to enable end-to-end provenance support on real HPC systems with little manual effort. The PROV-IO$^+$ framework can support both containerized and non-containerized workflows on different HPC platforms with  flexibility in selecting various classes of provenance. Our experiments with realistic workflows show that PROV-IO$^+$ can address the provenance needs of the domain scientists effectively with reasonable performance (e.g., less than 3.5\% tracking overhead for most experiments).
Moreover, PROV-IO$^+$ outperforms a state-of-the-art system (i.e., ProvLake) 
in our experiments.
\end{abstract}

}

\maketitle

\IEEEdisplaynontitleabstractindextext

%
\IEEEpeerreviewmaketitle

\section{Introduction}
\label{sec:intro}

\subsection{Motivation}
\label{sec:motivation}
Data-driven scientific discovery has been well acknowledged as a new fourth paradigm of scientific innovation~\cite{TonyHey-4thParadigm-MSR2009}. The shift toward the data-driven paradigm imposes new challenges in data findability, accessibility, interoperability,  reusability (i.e., FAIR principles~\cite{wilkinson2016fair,FAIR}) and trustworthiness~\cite{MartinMilton-TrustworthyData-Nature2020}, all of which demand innovative solutions for modeling and capturing provenance, i.e., the lineage of data life cycle.

\begin{figure}[htbp]
\centering
\centerline{\includegraphics[width=3in]{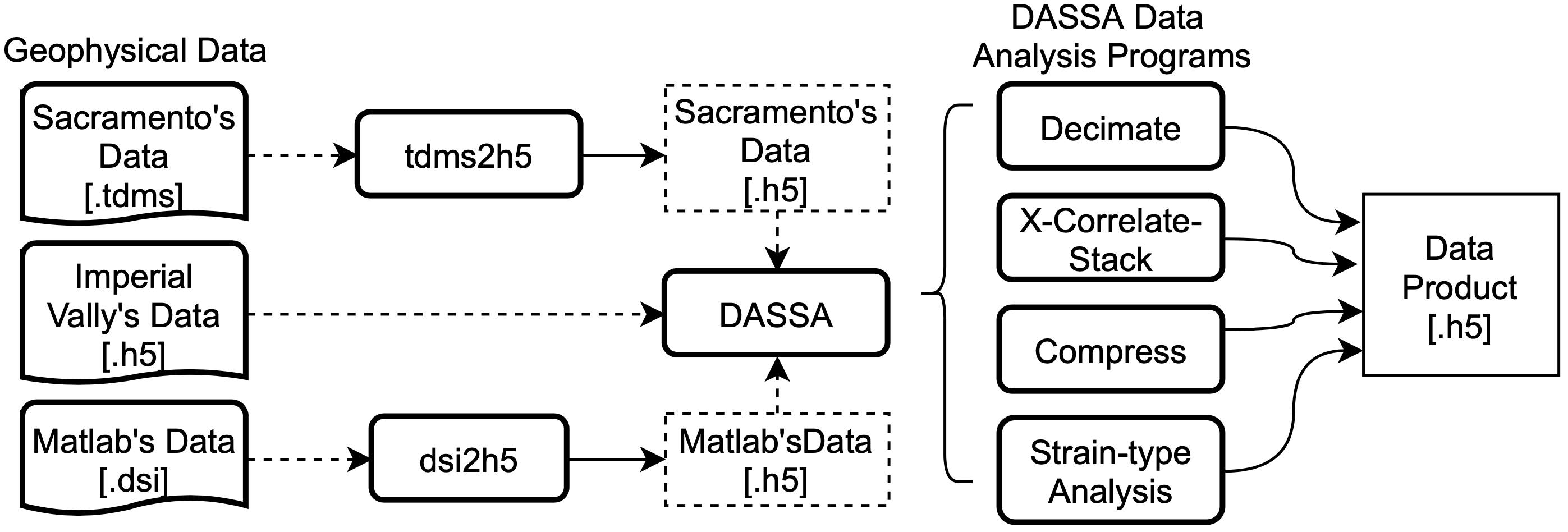}}
\vspace{-0.05in}
\caption{{\bf DASSA workflow.} Solid arrows stand for write operation and dashed arrows stand for read operation. 
}
\label{fig:DASSA}
\end{figure}

As an example, Figure~\ref{fig:DASSA} shows a simplified scientific workflow  which analyzes  geophysical sensing data on high performance computing (HPC) systems (i.e., DASSA~\cite{DASSA-paper}) .
The workflow takes geophysical data as  input, which are often stored in different file formats (e.g., ``.tdms'', ``.h5''). It then converts non-HDF5 files into  a uniform HDF5 format (i.e., ``.h5'').  
Depending on the analysis goals, the workflow further applies a set of  analysis programs (e.g., ``Decimate'', ``X-Correlation-Stacking'') to process the files, the results of which are stored as data products in HDF5 format.

Based on our survey, the domain scientists using DASSA need
the fine-grained origin of the data products (i.e., \textit{backward data lineage}). For example,   {\it User A} applies the  ``Decimate'' program with a number of HDF5 files as input  and generates a set of data products. Another {\it User B} may query the origin of the datasets in the final data products to understand which datasets in the input files contributed to which portions of the final data products, or who initiated the ``Decimate'' application  to generate the data products and when.
Such provenance information is important for ensuring the reproducibility, explainability, and security of the DASSA data. 

Nevertheless, the DASSA workflow involves multiple programs accessing multiple files
using different  I/O interfaces and operations (e.g., HDF5 and POSIX), which makes tracking and deriving the data provenance non-trivial. 
Moreover, as we will elaborate in Section~\S\ref{sec:casestudies}, there are other diverse needs of provenance for different scientific workflows and data (e.g., I/O statistics,  configuration lineage). Such diversity, complexity, as well as the stringent performance requirement in HPC environments call for a practical  solution beyond the state of the art. 

\vspace{-0.1in}
 \subsection{Limitations of State-of-the-art Tools}
\label{sec:limitations}
Unfortunately, to the best of our knowledge, existing provenance tools cannot address the grand challenge above sufficiently due to
a number of limitations:
 
First, while the importance of provenance has been well recognized across communities in general (e.g., databases~\cite{whyandwhere,TaPP15,ProvenanceandProbabilities,GoingBeyondProvenance,ProvenanceonParallelDB}, operating systems (OS) ~\cite{PASS,Margo-Layering-ATC09}, eScience~\cite{Silico,ASurveyine-science,challengesandopportunities,BigProvenanceStream,EfficientRuntimeCapture}), there is a lack of concrete understanding of the exact provenance needs of domain scientists, largely due to the variety of data and metadata that could be generated from HPC systems. As a result, existing solutions are often too coarse-grained (e.g., whole file tracking without understanding HPC data formats ~\cite{PASS}) to help domain scientists effectively, 
or too specific for one use case (e.g., Machine Learning~\cite{PROV-ML})  to support general needs.    

Second, in terms of provenance modeling, we find that existing standards (e.g., W3C PROV~\cite{PROV}) are too vague to describe the characteristics of scientific data provenance precisely. Scientists often seek a variety of information from scientific workflows on  HPC systems,  including the origins of data products, the configurations used for deriving results, the usage patterns of datasets,  and so on, which cannot be described effectively using any existing provenance models. Such ambiguity  limits the capability of existing provenance solutions for describing scientific data.

Third, in terms of usability, 
existing approaches often require the users to identify the critical code sites in the workflow software (e.g., loop structure~\cite{ProvLake}) and manually insert  API calls to track the desired information accordingly. 
Moreover, they often rely on many external packages to work properly, which make it difficult to deploy and use them  on different HPC platforms. 
The labor-intensive and error-prone approaches, together with the portability and compatibility issues,  hinder the wide adoption  of provenance products and diminishes the potential benefits. 

Note that the limitations highlighted  above are correlated. For example, the lack of understanding of provenance needs and the ambiguity of the provenance model are contributing to each other, which fundamentally limits the usability of existing solutions in terms of  granularity, expressibility, etc., which in turn makes clarifying the ambiguity and real needs difficult. 


\subsection{Key Insights \& Contributions}

We tackle the  grand challenge of  provenance support for scientific data on HPC systems in this paper. 

First, we observe that for a provenance framework to be practical and useful, inputs from the end users (i.e., domain scientists) is essential.
Therefore,  we collaborate with domain scientists to analyze four representative scientific workflows in depth. 
In doing so, we identify the unique characteristics of the workflows studied (e.g., I/O interfaces, data formats, access patterns) as well as the specific needs for scientific data provenance (e.g., lineage at file, dataset, or attribute granularity). 

Second, we observe that I/O operations are critically important in affecting the state of data that form the lineage needed by the domain scientists.
Therefore, different from existing solutions~\cite{ProvLake,Karma, OceanObservatory}, 
we introduce an  I/O-centric provenance model dedicated for the HPC environments.
The model enriches the W3C PROV standard~\cite{PROV} with a variety of concrete sub-classes, which can describe both the data and the associated I/O operations and execution environments precisely with extensibility. 
Moreover, it enables us to decouple the data provenance from specific executions of a workflow and support  the integration of provenance from multiple runs naturally,
which is important as workflows may evolve over time.



Third, based on the fine-grained provenance model, we find that the rich  I/O middleware 
already used by the scientists  provide an ideal vehicle for capturing the desired  provenance transparently. 
Therefore, we create a configurable and extensible library and integrate it with
 existing I/O code paths (e.g., HDF5 I/O and POSIX syscalls) 
to capture necessary information without requiring the scientists to 
modify the source code of their workflows.
Moreover, to further improve the usability, we persist the captured provenance 
 as standard RDF triples~\cite{rdf} and enable provenance query and visualization.
 

{Forth, through the communication with domain scientists in the industry,  we notice the increasing importance of supporting containerization~\cite{singularity}. By wrapping the HPC workflows together with their dependencies in containers,
the containerization techniques can effectively reduce the burden of software maintenance and thus enable more desired features including reproducibility, reusability, interoperability, etc.
Therefore, 
a provenance framework should be generic enough to  handle provenance in both containerized and non-containerized scenarios.
}

{Based on the key ideas above, we build a framework called PROV-IO$^+$, 
which can provide end-to-end provenance support for domain scientists 
with little manual effort across platforms.
We deploy PROV-IO$^+$ on  representative supercomputers and evaluate it with realistic workflows. Our experiments show that PROV-IO$^+$ incurs reasonable performance overhead
and outperforms a state-of-the-art provenance product (i.e., IBM ProvLake~\cite{ProvLake}) 
for the use cases evaluated. More importantly, through the query and visualization support, PROV-IO$^+$ can address the provenance needs of the  scientists effectively.}

In summary, we have made the following contributions:
{
\begin{itemize}
\item Identifying concrete provenance needs of domain scientists based on four representative scientific workflows;
\item Designing a comprehensive PROV-IO$^+$  model to describe the provenance of scientific data precisely and extensibly;
\item Building a practical prototype of  PROV-IO$^+$  which can support different HPC workflows with little human efforts in both containerized and non-containerized scenarios; 
\item Measuring the PROV-IO$^+$ prototype in HPC environments and demonstrating the efficiency and effectiveness; 
\item Releasing  PROV-IO$^+$ as an open-source tool to facilitate follow-up research on provenance in general.
\end{itemize} }


\subsection{Experimental Methodology \& Artifact Availability}
\label{sec:intro-exp}
{Experiments were performed on 
three state-of-the art platforms, including 
LBNL Cori Supercomputer~\cite{cori}, Samsung SAIT Supercomputer (SuperCom)~\cite{SAITSuperCom}, and Google Cloud Platform (GCP)~\cite{GCP}.
First, in terms of non-containerized scenario, we applied  PROV-IO$^+$ to three scientific workflows (i.e., DASSA~\cite{DASSA-paper}, Top Reco~\cite{TopReco}, and an I/O-intensive application based on H5bench~\cite{h5bench}) on the Cori supercomputer. Second, in terms of containerized scenario, we applied  PROV-IO$^+$ to analyze one deep learning workflow (i.e., Megatron-LM~\cite{Megatron1}) on SuperCom. 
As will be discussed  in \S\ref{sec:casestudies}, these use cases cover diverse characteristics  (e.g.,  various languages, file formats, I/O interfaces, metadata) and provenance needs (e.g., file/dataset/attribute lineage,  metadata versioning, I/O statistics). 
%
%
We varied the critical parameters of the workflows 
to measure the run-time performance and storage requirements under a wide range of scenarios. 
Third, we examined container's impact on provenance tracking with  Megatron-LM~\cite{Megatron1} on the Google Cloud Platform where we can schedule both containerized and non-containerized workflows and perform a fair comparision.
In addition, 
we compared PROV-IO$^+$ with ProvLake~\cite{ProvLake} using the Python-based Top Reco workflow as ProvLake only supports Python at the time of this writing.
The PROV-IO$^+$ prototype is open-source at  \url{https://github.com/hpc-io/prov-io}.}

\section{Background}
\label{sec:bg}

\subsection{W3C Provenance Standard}
\label{sec:w3c}

\begin{figure}[htbp]
\centering
\centerline{\includegraphics[width=2.8in]{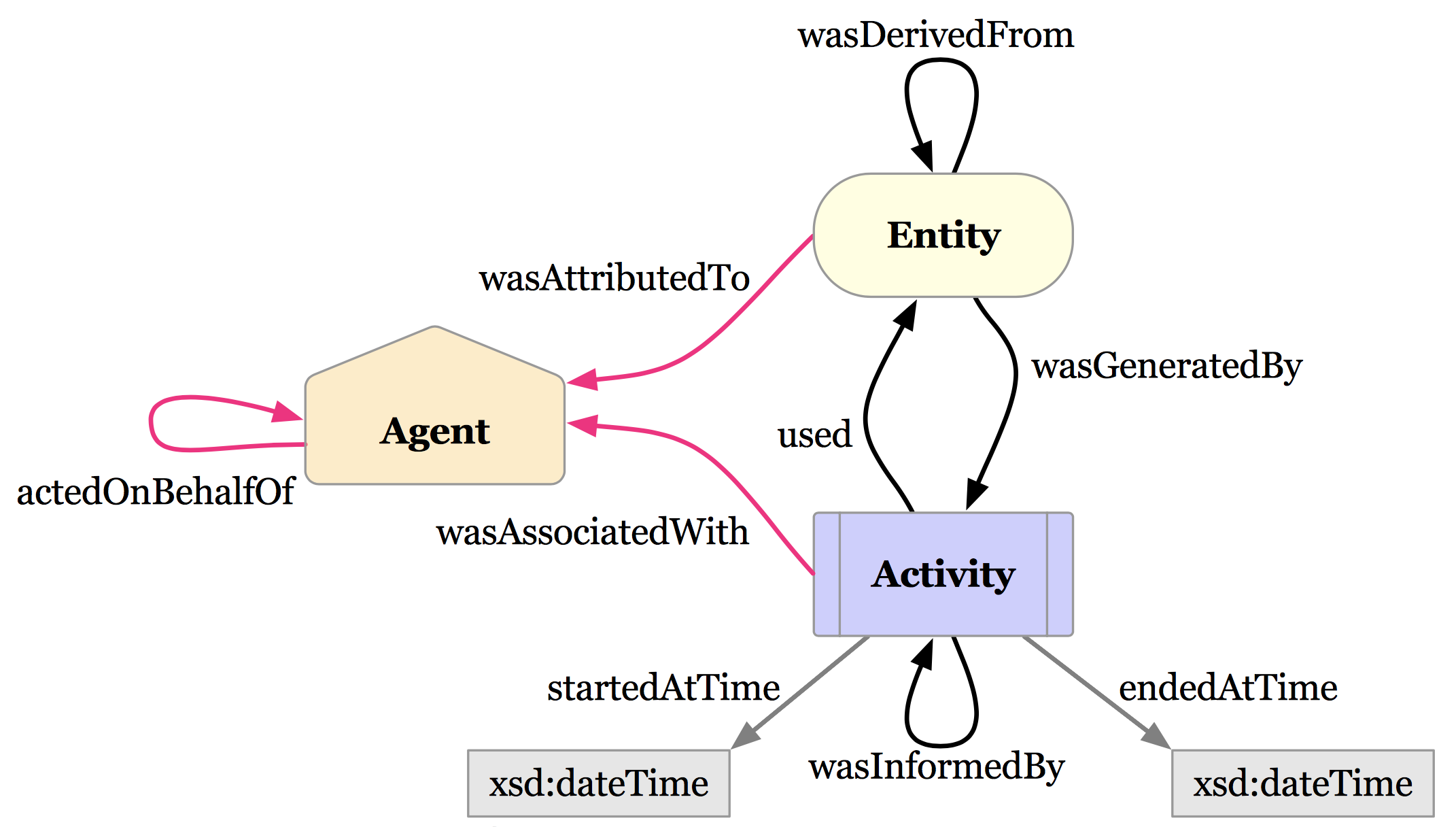}}
\caption{The W3C Provenance Model ~\cite{TheW3CPROVfamily}.}
\label{fig:PROV-DM}
\end{figure}

The PROV family of specifications, published by the World Wide Web Consortium (W3C), is a set of provenance standard to promote provenance publication on the Web with interoperability across diverse provenance management systems~\cite{TheW3CPROVfamily}. 
One key specification is PROV-DM, an extensible relational model which describes provenance information with a graph representation. 
As shown in Figure~\ref{fig:PROV-DM},
a W3C provenance graph abstracts information into classes of {\it Entity}, {\it Activity}, {\it Agent}, and {\it Relation} between the first three classes. 
Another critical specification is PROV-O which describes the mapping of PROV-DM classes to RDF triples. In PROV-O, {\it Entity}, {\it Activity} and {\it Agent} are mapped to subjects and objects, while {\it Relation} is mapped to predicates. 
We follow the W3C PROV standard in the PROV-IO$^+$ design.

\subsection{HPC I/O Libraries}

I/O libraries (e.g., ADIOS~\cite{ADIOS}, HDF5~\cite{hdf5}, and NetCDF~\cite{NetCDF}) play an essential role in scientific computations. Many workflows leverage the library I/O to manipulate data files. For example, HDF5 (i.e., Hierarchical Data Format version 5) is one of the the most wildly used I/O libraries for scientific data~\cite{HDF5-Usage}. It is developed to be a parallel data management middleware to bridge the gap between HPC applications and the complicated, low-level details of underlying file systems, and has grown to a popular data format and management system.

In this work, we integrate our solution with the HDF5 library 
besides the classic POSIX I/O operations. This is based on the observation that 
HDF5 has evolved with a Virtual Object Layer (VOL) which can intercepts object-level API operations to functional plugins, called VOL Connectors~\cite{vol-provenance}. VOL connectors allow third-party developers to add desired storage functionalities, which can be loaded dynamically at runtime. 
We leverage such extensibility for tracking the provenance of HDF5 I/O data.


\section{Case Studies}
\label{sec:casestudies}

\begin{table*}[h!]
    \centering
    \caption{Four Real Use Cases with Different Characteristics and Provenance Needs.}
        \vspace{-0.1in}
    \begin{adjustbox}{max width=\textwidth}
    \begin{tabular}{c|c|c|c}
    \hline
    {\bf Use Case}    & {\bf Description} & {\bf I/O Interface} & {\bf Provenance Need}\\
           \hline
    Top Reco & {\small training GNN models for top quark reconstruction; multi-program, multi-file}; &  POSIX & {\small metadata version control \& mapping}  \\
    DASSA & {\small parallel processing of acoustic sensing data;  multi-program, multi-file;}  & HDF5 \& POSIX  & {\small backward lineage of data products}\\
    H5bench & {\small simulating typical I/O patterns of HDF5 app; multi-program, single-file; }& HDF5  & {\small I/O statistics \& bottleneck} \\
    Megatron-LM & {\small parallel transformer model for NLP; multi-program, multi-file; }& POSIX  & {\small Checkpoint-configuration consistency} \\
        \hline
    \end{tabular}
    \end{adjustbox}
    \label{tab:workflows_introduction}
    \vspace{-12pt}
\end{table*}

In this section, we discuss four real-world use cases to motivate the  I/O-centric provenance further.
For each case, we describe its semantics and characteristics, 
the provenance need of the domain scientists, 
and the associated challenges. 

\vspace{-8pt}
\subsection{Top Reco - Lineage of configurations}
\label{sec:topreco}

\noindent
{\bf Workflow Description.} Top Reco~\cite{TopReco} is a Machine Learning (ML) workflow in high-energy physics data analysis, which uses Graph Neural Network (GNN) models
for top quark reconstruction. 
Top quarks are the elementary particles with the most mass that may decay quickly and are not detectable directly due to their mass.
By representing particles and their relationships as graphs, the GNN-based workflow can help reconstruct top quarks more accurately and efficiently,
which is important for physics discoveries.

\begin{figure}[htbp]
\centering
\centerline{\includegraphics[width=3in]{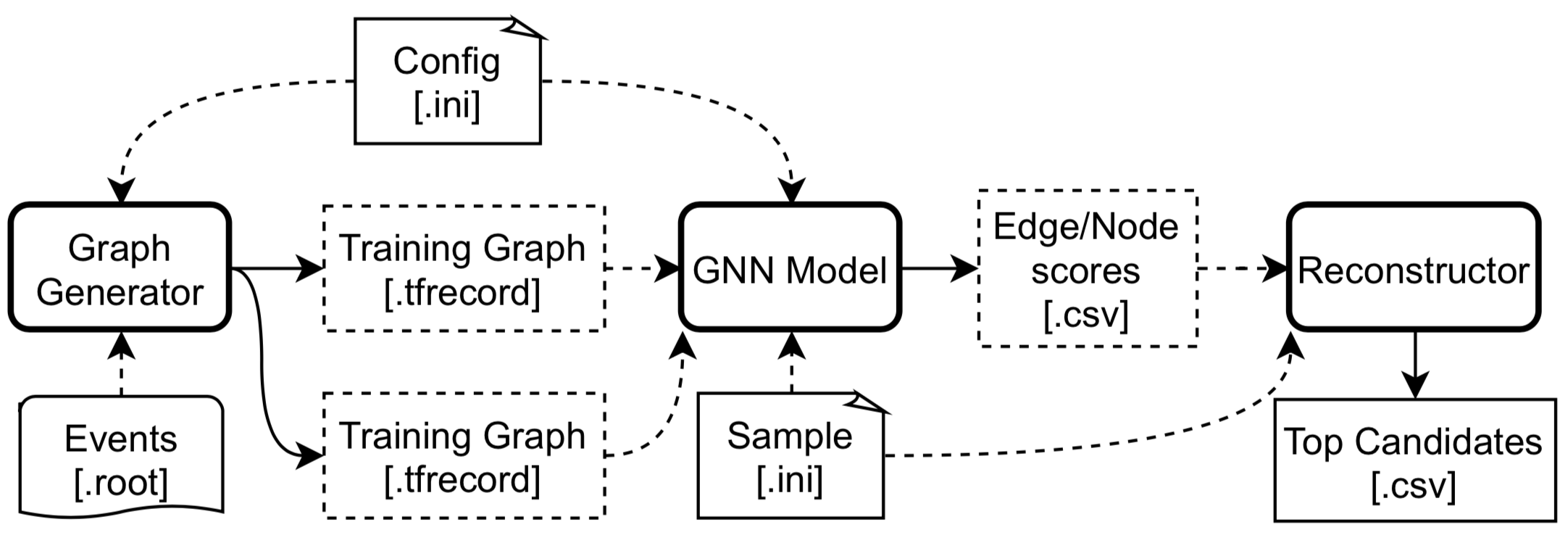}}
    \vspace{-0.1in}
\caption{{\bf Top Reco workflow.} Solid arrows stand for write operation and dashed arrows stand for read operation.}
\label{fig:TopReco}
\end{figure}

In Figure~\ref{fig:TopReco}, we show the key steps of the Top Reco workflow. First, the workflow takes two types of  files as input, including the ``.root'' file for input event and the ``.ini'' file for configuration. 
Second, it generates ``.tfrecord'' files which stores the training dataset and test dataset based on the input events. Third, it trains a GNN model with the training dataset and tests the model with the test dataset by accessing the ``.tfrecord'' files.
Fourth, a range of scores of edge and nodes are generated as the output of the model.
Finally, a reconstructor component runs a simulation of reconstructing the top quarks based on highest  scores.
As summarized in  Table~\ref{tab:workflows_introduction}, the Top Reco workflow uses the POSIX I/O interface, and involves multiple programs accessing multiple files.

\smallskip
\noindent
{\bf Provenance Need.} 
In the Top Reco case, the domain scientists are interested in the impact of GNN configurations on the model performance. 
Specifically, they would like to know which combination of model hyperparameters and dataset preselections result in the best training accuracy. 
In other words, they would like to have fine-grained version control of the metadata (e.g., hyperparameters, preselections) 
as well as the correlation between the metadata and the  result to ensure the explainability and  reproducibility.

\smallskip
\noindent
{\bf Challenges.} 
 Essentially, the Top Reco case requires automatic version control management on the machine learning model. 
 However,  a typical version control system (e.g., Git) cannot meet the requirements because it cannot automatically track the model performance and maps the performance to the model configuration. 
In practice, the scientists may need to execute the workflow for multiple times with different configurations, and each execution may  take multiple hours or more.  Due to lack of provenance support, 
the scientists  have to manually make a new copy of configuration when 
they start a new run,
and record  the corresponding result later.
 Such common practice is time-consuming and not scalable. In other words, a new provenance framework is urgently needed.

\vspace{-8pt}
\subsection{DASSA - Lineage of Data Products}
\label{sec:dassa}
\noindent
{\bf Workflow Description.} 
As mentioned in Section~\S\ref{sec:motivation}, 
DASSA~\cite{DASSA-paper} is a parallel storage and analysis framework for distributed acoustic sensing (DAS). It uses a hybrid (i.e., MPI and OpenMP) data analysis execution engine to 
support efficient and automated parallel processing of geophysical data in HPC environments, which has been  applied for accelerating a variety of scientific computations including earthquake detection, environmental characterization, and so on.
The overall workflow is described in Figure~\ref{fig:DASSA}.




\smallskip
\noindent
{\bf Provenance Need.} As discussed in Section~\S\ref{sec:motivation}, 
the domain scientists need the \textit{backward data lineage} to understand the origin of  the data products and to ensure the data reproducibility, explainability, and security, among others.

\smallskip
\noindent
{\bf Challenges.} 
The DASSA workflow may involve multiple different programs, file formats, I/O interfaces, and end users, which is representative for large-scale scientific workflows in  HPC environments.
Moreover, both 
the file level and the sub-file level (e.g., inner hierarchies of the HDF5 format) information is needed.
To the best of our knowledge, 
none of the existing  provenance models or systems can handle the complexity to meet the comprehensive needs. 



\vspace{-8pt}
\subsection{H5bench - Data usage and I/O performance}
\label{sec:h5bench}

\noindent
{\bf Workflow Description.}
H5bench~\cite{h5bench} is a parallel I/O benchmark suite for HDF5~\cite{Folk:2011:OHT} that is representative of various large-scale workflows. It includes a default set of read and write workloads with typical I/O patterns in HDF5 applications on HPC systems,
which enables creating synthetic workflows  to simulate diverse HDF5 I/O operations in HPC environments. The benchmark also contains ‘overwrite’ and ‘append’ operations that allow modifying data or metadata of existing files and appending new data, respectively. 
We collect an H5bench-based workflow which contains a
combination of  ‘write’, ‘overwrite’, 'append' and ‘read’ workloads operating on HDF5 files via MPI. This workflow simulates the typical scenarios where a single file may be accessed concurrently by HPC applications and multiple versions of a dataset may be generated accordingly.
As shown in  Table~\ref{tab:workflows_introduction}, the H5bench-based workflow   mainly uses the  HDF5 I/O interface, and involves multiple programs accessing a single file.



\smallskip
\noindent
{\bf Provenance Need.} 
Understanding frequently accessed data in large datasets leads to optimizing I/O performance by improved data placement and layout.
Scientists typically use the H5bench-based workflow to collect I/O statistics and identify potential bottlenecks on HPC systems.
While I/O profiling tools, such as Darshan \cite{Darshan} and Recorder \cite{Recorder-pdsw} collect coarse-grained statistics of I/O performance, there are no tools to extract data access information and the cost of those operations. 
Fine-grained information such as the
total number of each type of HDF5 I/O operations incurred during the workflow, 
the accumulated time cost for each type of operations, 
the distribution of operations and time overhead, 
the HDF5 APIs invoked at a specific time point, etc. 
would be critically important for understanding the system behavior and fine-tuning the performance.


\smallskip
\noindent
{\bf Challenges.} 
The H5bench use case involves handling HDF5 datasets  concurrently and measuring diverse fine-grained metrics at the HDF5 API level,
which requires deep understanding of the semantics and internals of HDF5. 
Since existing solutions are largely incompatible with HDF5, 
they are fundamentally inapplicable for this important category of use cases.




\vspace{-6pt}

\subsection{Megatron-LM - Checkpoint Consistency}
\label{sec:megatron-lm}

\begin{figure}[htbp]
\centering
\centerline{\includegraphics[width=3in]{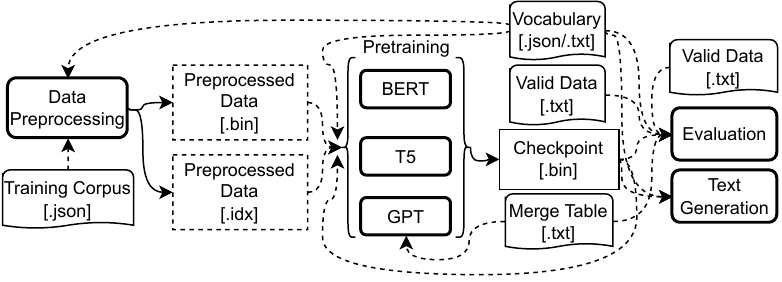}}
\caption{{\bf Megatron-LM Workflow.}  Solid arrows stand for write operation and dashed arrows stand for read operation.}
\label{fig:Megatron-LM}
\end{figure}

\noindent
{\bf Workflow Description.} 
{
Megatron-LM is based on Megatron, which is a powerful transformer (i.e., a type of deep learning models) developed by  NVIDIA~\cite{Megatron1,Megatron2,Megatron3}.
Megatron supports training large transformer language models at scale, which is achieved by providing efficient, model-parallel (tensor, sequence, and pipeline), and multi-node pre-training of transformer-based models (e.g., GPT~\cite{GPT}, BERT~\cite{BERT}, and T5~\cite{T5}) using mixed precision. 
Megatron-LM  scales the transformer training by supporting data parallelism and model parallelism further. Specifically,
the data parallelism  is achieved by splitting the input dataset across  specified devices (e.g., GPUs); on the other hand, the model parallelism is implemented by splitting the execution of a single transformer module over multiple GPUs working on the same dataset.
Both data parallelism and model parallelism features can be optionally configured in the workflow, and they are both enabled in this study for completeness.
}
{
Figure~\ref{fig:Megatron-LM} shows a simplified overview of the Megatron-LM workflow. First, a training corpus (``.json'') is preprocessed by the data processing module, which generates  a binary file  (``.bin'') and an index file (``.idx''). 
The preprocessed data  become the input of the pretraining transformer models. A trained model (i.e., checkpoint) will be generated at the end of pretraining, and it can be used in the follow-up evaluation or text generation. Users may also skip the pretraining step if they already have a trained model available. 
}

\smallskip
\noindent
{\bf Provenance Need.} 
{
In the Megatron-LM workflow, the consistency between pretraining models's checkpoint and the corresponding configuration is important. The checkpoint mainly records the metadata of the previous pretraining process, such as micro/global batch size and state of optimizer/scheduler, which are dependent on the configuration parameters. Blindly modifying configuration parameters in the next pretraining could easily result in various types of errors (e.g., network errors, test code errors). Moreover, many  configuration parameters are tightly correlated with each other, changing configuration parameters without preserving the correlation may also  lead to failures. In addition, adjusting pretraining hyperparameters incautiously may  affect the model quality  negatively (e.g., result in overfitting). Therefore, in the Megatron-LM use case, the domain scientists want to track the checkpoint and configuration provenance to ensure the checkpoint-configuration consistency.}

\smallskip
\noindent
{\bf Challenges.} 
{
The challenge for the Megatron-LM workflow is two-fold.
First of all,
the workflow involves hundreds of configuration entries and checkpointed statuses which are difficult to track or reason by human.
Due to the lack of tool support, the domain scientists  cannot  manage checkpoints generated by multiple training processes conveniently and identify a qualified checkpoint and the associated configurations consistent with  new training processes.}

{
Moreover, with the growing popularity of container technologies~\cite{Docker, k8s}, HPC systems have started to be integrated with container-based job runtime tools~\cite{singularity}. 
In such HPC systems, large-scaled parallel pretraining models are executed in a containerized environment, which largely avoids tedious efforts in resolving installation or runtime dependencies (e.g, PyTorch~\cite{pytorch} and nccl~\cite{nccl}). 
Besides the challenge of the Megatron-LM workflow itself, the domain scientists using the workflow would like to execute the workflow in a containerized HPC environment. 
Since a container is an isolated environment by design where application cannot directly interactive with the host system, how to containerized Megatron-LM and track provenance information of the containerized workflow at scale on HPC systems is another major challenge in this use case.}

{
Note that both Megatron-LM and Top Reco (~\S\ref{sec:topreco}) belong to deep learning provenance use cases. However, Megatron-LM has two major differences compared to Top Reco. First, in Megatron-LM, the provenance need is checkpoint-configuration consistency, while Top Reco's provenance need is configuration version control. Second, Top Reco is a traditional single thread workflow, while Megatron-LM is a containerized parallel workflow, which introduces more challenges in terms of both provenance tracking and provenance storage.
}

\vspace{-0.05in}
\subsection{Summary}
\label{sec:case_study_summary}

By analyzing the {four} cases in depth and consulting with the domain scientists, we find that 
there is a big gap between the provenance needs and  existing  solutions. The variety of the workflow characteristics 
(e.g., different I/O interfaces and file formats) as well as the diversity of scientists' needs 
motivates us to design a comprehensive provenance framework to address the  challenge, which we elaborate in the following sections.

\section{PROV-IO$^+$ Design}
\label{sec:design}
In this section, we introduce the design of PROV-IO$^+$. We focus on the provenance model (\S\ref{sec:model}) and its system architecture (\S\ref{sec:architecture}), which are two fundamental pillars of PROV-IO$^+$. We defer additional implementation details to the next section (\S\ref{sec:implementation}).

\begin{figure*}[htbp]
\centering
\subfloat[PROV-IO$^+$ model.]{\includegraphics[width=4.9in]{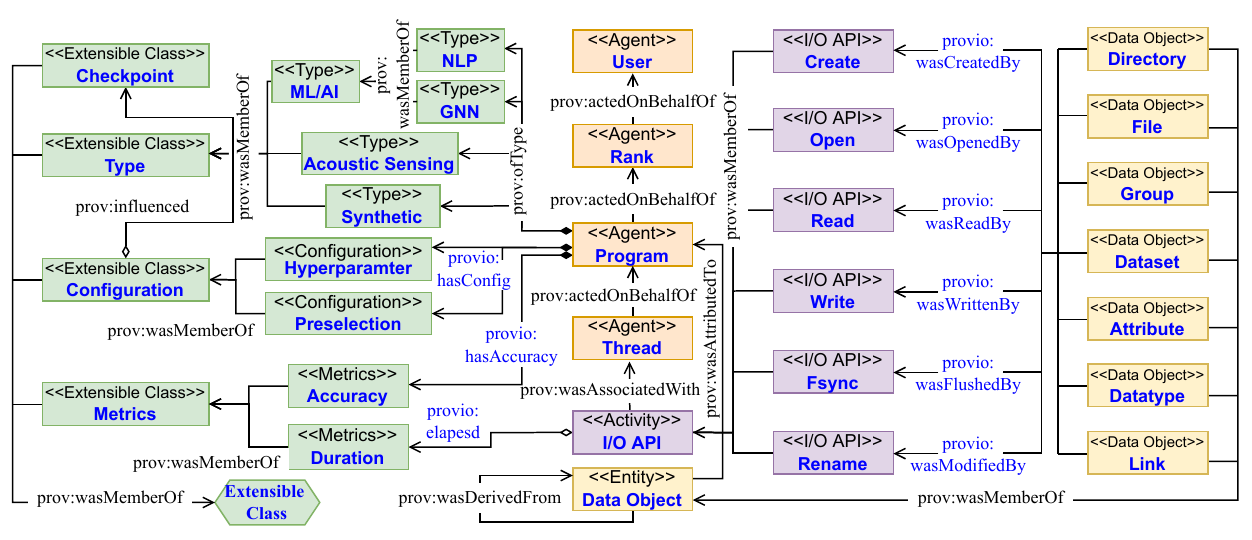}%
\label{fig:model}}
\hfil
\subfloat[A Provenance Snippet.]{\includegraphics[width=2.1in]{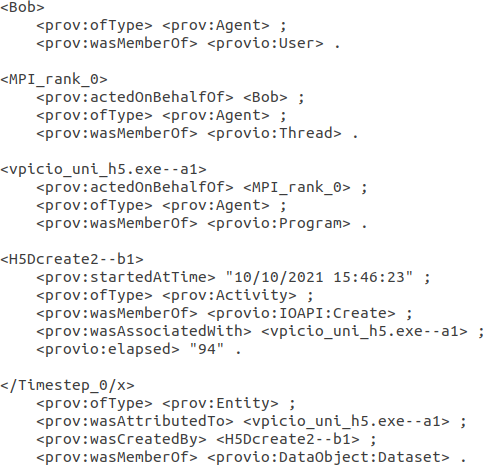}%
\label{fig:model_example}}
\label{fig_sim}
    \vspace{-0.1in}
\caption{{\bf (a) PROV-IO$^+$ Model Overview}. The PROV-IO$^+$ model classifies information into five super-classes: {\it Entity} (yellow boxes), {\it Activity} (purple boxes), {\it Agent} (orange boxes), {\it Extensible Class} (green boxes) and {\it Relation} (text on arrows). 
The new concepts introduced by PROV-IO$^+$ are highlighted with blue color. (b) A Provenance Snippet based on PROV-IO$^+$.
}
\label{PROV-IO$^+$}
\end{figure*}

\subsection{PROV-IO$^+$ Model}
\label{sec:model}
Figure~\ref{PROV-IO$^+$}(a) shows an overview of the PROV-IO$^+$ model, which is derived based on the W3C  standard (\S\ref{sec:w3c}) as well as  the characteristics of typical workflows and the provenance needs of domain scientists (\S\ref{sec:casestudies}).

Following the W3C specification, we classify information into five PROV-IO$^+$ super-classes: {\it Entity} (yellow boxes in Figure~\ref{PROV-IO$^+$}(a)), {\it Activity} (purple boxes), {\it Agent} (orange boxes), {\it Extensible Class} (green boxes) and {\it Relation} (text on arrows).    
Moreover, we introduce a variety of concrete sub-classes to enrich the model, which can capture the data with different granularity as well as the associated I/O operations and execution environments  for deriving the data.
We summarize the definitions of the sub-classes in Table~\ref{tab:PROV-IO$^+$}
and highlight the main concepts added to each super-class as follows: 

\begin{table}[h]
    \centering
    \caption{Description of PROV-IO$^+$ Model.}
    \vspace{-0.1in}
    \begin{tabular}{m{1cm} | p{1.9cm} p{5.1cm}}
        \hline
        {\bf Super-class} & {\bf Sub-class} & {\bf Description} \\
        \hline
        & \cellcolor{Gray} {\footnotesize$<<$Data Object$>>$} \newline   Directory & \cellcolor{Gray} POSIX file system directory. \\ 
        & {\footnotesize$<<$Data Object$>>$} File & POSIX file system file.\\
        & \cellcolor{Gray} {\footnotesize$<<$Data Object$>>$} Group & \cellcolor{Gray} I/O library interior group structure (e.g., HDF5 group). \\
        \it{Entity} & {\footnotesize$<<$Data Object$>>$} Dataset & I/O library interior dataset structure (e.g. , HDF5 dataset). \\
        & \cellcolor{Gray} {\footnotesize$<<$Data Object$>>$} Attribute & \cellcolor{Gray} POSIX Inode extended attribute and I/O library interior attribute structure (e.g., HDF5 attribute). \\
        & {\footnotesize$<<$Data Object$>>$} Datatype & I/O library interior datatype structure (e.g., HDF5 datatype). \\
        & \cellcolor{Gray} {\footnotesize$<<$Data Object$>>$} Link & \cellcolor{Gray} POSIX file system hard/soft link. \\
        \hline
        &  {\footnotesize$<<$I/O API$>>$} Create & POSIX syscall ``open'' and I/O library ``Create'' APIs (e.g., H5Acreate). \\
        & \cellcolor{Gray} {\footnotesize$<<$I/O API$>>$} Open & \cellcolor{Gray} I/O library ``Open'' APIs (e.g., H5Aopen). \\
        & {\footnotesize$<<$I/O API$>>$} Read & POSIX syscall ``read'' (and variants) and I/O library ``Read'' APIs (e.g., H5Aread). \\
        \it{Activity} & \cellcolor{Gray} {\footnotesize$<<$I/O API$>>$} Write & \cellcolor{Gray} POSIX syscall ``write'' (and variants) and I/O library ``Write'' APIs (e.g., H5Awrite). \\
        & {\footnotesize$<<$I/O API$>>$} Fsync & POSIX syscall ``fsync'' (and variants) and I/O library ``Flush'' APIs (e.g., H5Flush). \\
        & \cellcolor{Gray} {\footnotesize$<<$I/O API$>>$} \newline Rename & \cellcolor{Gray} POSIX syscall ``rename'' (and variants) and I/O library  ``Rename'' APIs. \\
        \hline
        & User & Workflow user. \\
        \it{Agent} & \cellcolor{Gray} Rank & \cellcolor{Gray}  Individual MPI rank. \\
        &  Program &  Program instance. \\
        & \cellcolor{Gray} Thread & \cellcolor{Gray}  A thread of a program. \\
        \hline
        &  Checkpoint & 
        Checkpoint information of a program/workflow (e,g, checkpoint path, checkpoint status). \\
        {\footnotesize\it{Extensible Class}} &  \cellcolor{Gray} Type &
        \cellcolor{Gray} Type of a program/workflow (e.g., Machine Learning (Top Reco), Acoustic Sensing (DASSA)). \\
        &  Configuration & 
        Workflow configurations (e.g., hyperparameter in Top Reco).  \\
        & \cellcolor{Gray} Metrics & 
        \cellcolor{Gray} Evaluation metrics of the workflow. E.g., model accuracy in Top Reco. \\
        \hline
        &  provio: \newline wasCreatedBy & The relation between a $<<$I/O API$>>$ Create and a $<<$Data Object$>>$. \\
        & \cellcolor{Gray} provio:\newline wasOpenedBy & \cellcolor{Gray} The relation between a $<<$I/O API$>>$ Open and a $<<$Data Object$>>$. \\
        \it{Relation} & provio:\newline wasReadBy & The relation between a $<<$I/O API$>>$ Read and a $<<$Data Object$>>$. \\
        & \cellcolor{Gray} provio:\newline wasWrittenBy & \cellcolor{Gray} The relation between a $<<$I/O API$>>$ Write and a $<<$Data Object$>>$. \\
        & provio:\newline wasFlushedBy & The relation between a $<<$I/O API$>>$ Fsync and a $<<$Data Object$>>$. \\
        & \cellcolor{Gray} provio:\newline wasModifiedBy & \cellcolor{Gray} The relation between a $<<$I/O API$>>$ Rename and a $<<$Data Object$>>$. \\
        \hline
    \end{tabular}
    \label{tab:PROV-IO$^+$}
\end{table}




\subsubsection{Entity}
This PROV-IO$^+$ super-class includes seven specific {\it Data Object} sub-classes (i.e., {\it Directory}, {\it File}, {\it Group}, {\it Dataset}, {\it Attribute}, {\it Datatype}, {\it Link}). 
Together, these sub-classes cover common
I/O structures and file formats. 
For example,  {\it  Attribute} is a combined sub-class that can map to both the HDF5 attributes and the extended attributes of an inode in a POSIX-compliant Ext4 file system~\cite{Ext4}. 

\subsubsection{Activity}
This super-class includes six specific {\it I/O API} sub-classes (i.e., {\it Create}, {\it Open}, {\it Read}, {\it Write}, {\it Fsync}, {\it Rename}).
These  sub-classes cover a wide range of commonly used I/O operations in HPC environments.
For example, {\it  Read} can map to HDF5 read-family operations (e.g., ``H5Gread'', ``H5Dread'', ``H5Aread'', ``H5Tread'') and POSIX system call ``read'' and its variants. Note that these  operations are applicable to other I/O libraries too (e.g., NetCDF~\cite{NetCDF}). 

\subsubsection{Agent}
This super-class includes a set of sub-classes representing the operator of a series of activities,
such as {\it Thread}, {\it User}, {\it Rank}, and {\it Program}.  {This fine-grained representation is necessary because HPC applications  are typically multi-threaded and are executed in parallel (e.g., a group of MPI processes with different ranks running on a cluter of nodes).} 


\subsubsection{Extensible class}
This super-class contains  properties pertained by entities, activities and agents. It is designed to be extensible because valuable information is often workflow-specific. {In the current prototype, we define four generic sub-classes (i.e., {\it Checkpoint}, {\it Type}, {\it Configuration}, {\it Metrics}) to cover a variety of valuable information  that cannot be described precisely in the native W3C specification (e.g., hyperparameters of ML models, checkpoints of AI model training).}

\subsubsection{Relation}
This super-class describes the diverse relations among other classes. We inherit the basic 
W3C provenance relations between 
\textit{entity} \& \textit{entity} ({prov:wasDerivedFrom}), 
\textit{entity} \& \textit{agent} ({prov:wasAttributedTo}), \textit{activity} \& \textit{agent} ({prov:AssociatedWith}), \textit{agent} \& \textit{agent} ({prov:actedOnBehalfOf}). Moreover, we introduce new relations between \textit{entity} \& \textit{activity} to precisely describe the relations between various I/O API and Data Object sub-classes
(e.g., {provio:wasCreatedBy}, provio:wasReadBy, provio:wasWrittenBy, provio:wasModifiedBy). 

To make the description more concrete, we show an example snippet of provenance captured by PROV-IO$^+$ in Figure~\ref{PROV-IO$^+$}(b).
The provenance snippet contains five records pertained by different subjects. Each subject can be an {\it Agent} (e.g., ``Bob'', ``MPI\_rank\_0''), an {\it Activity} (e.g., ``H5Dcreate2--b1''), or an {\it Entity} (e.g., ``/Timestep\_0/x''). Each record is a series of triples starting with a unique subject, where the triples describe provenance information of a subject. Note that the record length may vary depending on the provenance information associated with the subject. Given this  snippet, we can derive complex  provenance information (e.g., dataset ``/Timestep\_0/x'' was created by I/O API ``H5Dcreate2--b1'' associated with program ``vpicio\_un\_h5.exe--a1'' on thread ``MPI\_rank\_0'', which was started by user ``Bob'').


\subsection{PROV-IO$^+$ Architecture}
\label{sec:architecture}
\begin{figure*}[htbp]
\centering
\subfloat[Mode \#1: Provenance  for Non-Containerized Workflow.]{\includegraphics[width=2.9in]{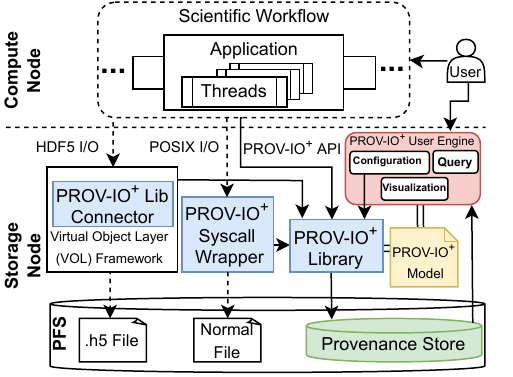}%
\label{fig:arch_traditional}}
\hfil
\subfloat[Mode \#2: Provenance  for Containerized Workflow.]{\includegraphics[width=2.9in]{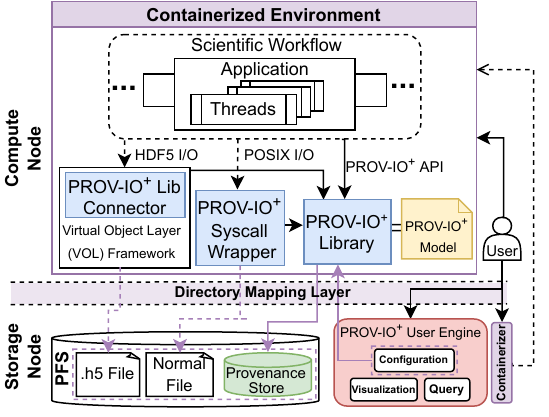}%
\label{fig:arch_containerized}}
\label{fig_sim}
\vspace{-4pt}
\caption{{\bf The Architecture of PROV-IO$^+$ Framework.} There are two usage modes: (a) Mode \#1 provides provenance support for non-containerized workflows; (b) Mode \#2 supports containerized workflows.  The  framework includes five major components in total: the PROV-IO$^+$ model (yellow), a provenance tracking engine (blue modules), a provenance store (green), a user engine (red), and a containerizer engine (purple). }
\label{architecture}
\end{figure*}

Figure~\ref{architecture} shows the architecture of the PROV-IO$^+$ framework, which supports two usage modes: Mode \#1 (Figure~\ref{fig:arch_traditional}) provides provenance support for traditional non-containerized workflows on HPC systems; Mode \#2 (Figure~\ref{fig:arch_containerized}) supports containerized workflows. 
There are five components in total, including:
(1) the PROV-IO$^+$ model (yellow) to specify the provenance information \S\ref{sec:model};
(2) a provenance tracking engine (blue modules) which captures I/O operations from multiple I/O interfaces;
(3) a provenance store (green) which persists captured provenance into RDF triples; 
(4) a user engine (red) for users to query and visualize provenance information; 
(5) a containerizer engine (purple) to support other components in  containerized  environments. 

Among the five components, the PROV-IO$^+$ model (yellow) has been discussed in details in \S\ref{sec:model}. We introduce the other three common components used in both modes (i.e.,  provenance tracking, provenance store, and user engine) one by one in \S\ref{sec:common-components}, and then discuss the containerizer engine for supporting containerized workflows in \S\ref{sec:containerizer}.

\subsubsection{\textbf{Mode \#1:  Support for Classic Workflows}}
\label{sec:common-components}

To  provide provenance support for the classic, non-containerized workflows (i.e., Mode \#1),  
PROV-IO$^+$ leverages three major components based on its provenance model (\S\ref{sec:model}) as follows:

\smallskip
\noindent
{\bf Provenance Tracking.}
\label{sec:tracking}
As shown in Figure~\ref{fig:arch_traditional}, a scientific workflow is typically started on compute nodes. 
The workflow may consist of several parallel applications with multiple threads running concurrently. During the workflow execution, all I/O operations (e.g., POSIX and HDF5) 
are monitored by PROV-IO$^+$ 
for provenance collection. 

Specifically, the Provenance Tracking component contains two thin modules (i.e., {\it PROV-IO$^+$ Lib Connector} and {\it PROV-IO$^+$ Syscall Wrapper}) for monitoring the library I/O 
and POSIX I/O operations respectively. 
In case of the HDF5 library, the PROV-IO$^+$ Lib Connector monitors the I/O requests within the HDF5 Virtual Object Layer (VOL). 
In case of POSIX, the I/O syscalls are monitored through the PROV-IO$^+$ Syscall Wrapper which is configurable via environmental variables.
In both cases, PROV-IO$^+$ let the native I/O requests pass through and invoke the core {\it PROV-IO$^+$ Library} for
collecting the provenance  defined by the PROV-IO$^+$ model without changing the original I/O semantics.  

Note that both the library I/O and POSIX I/O operations can be tracked in a transparent and non-intrusive way from the workflow's perspective, which is important for usability. 
  In addition, to achieve extensibility, 
  we provide a set of PROV-IO$^+$ APIs which enables users to convey user/workflow-specific semantics  and requirements to PROV-IO$^+$ (i.e., Extensible Class in PROV-IO$^+$ model).
Similar to ProvLake~\cite{ProvLake}, users can instrument their workflows with PROV-IO$^+$ APIs as needed (e.g., tracking a specific hyperparameter of a ML workflow).   
By providing such flexibility, additional  provenance needs can be satisfied by PROV-IO$^+$ conveniently.


\smallskip
\noindent
{\bf Provenance Store.}
\label{sec:storage}
The Provenance Store component maintains the provenance information as RDF graphs durably on the underlying parallel file system to enable future queries. 
We choose an RDF triplestore instead of a traditional SQL database
for two main reasons: (1) W3C PROV-DM already has a well-defined ontology (i.e., PROV-O\cite{PROV}) to map the model to RDF, so using RDF  makes PROV-IO$^+$ compatible with other W3C-compliant solutions; (2) To answer path queries in provenance use cases, SQL queries with repeated self-joins are necessary to compute the transitive closure, which often leads to worse performance when the provenance grows~\cite{SPADE-ProvenanceAuditing-Middleware12}.


More specifically, the Provenance Store component provides an interface for the PROV-IO$^+$ Library to manipulate provenance records and maintain  provenance graphs efficiently, which includes creating a new provenance RDF graph in memory, loading  an existing  graph,  inserting new records to an existing graph, etc.
To minimize the performance impact on  the workflow, 
the in-memory provenance graph is serialized to the Provenance Store asynchronously.
And depending on the need of the user, the serialization operation may be triggered either periodically  or by the end of the workflow.

\smallskip
\noindent
{\bf PROV-IO$^+$ User Engine.}
\label{sec:engine}
The provenance information could be enormous due to the complexity of scientific workflows.
To avoid distraction and help users 
derive insights, 
the PROV-IO$^+$ User Engine component allows users to 
enable/disable individual sub-classes defined in the PROV-IO$^+$ model, 
which also enables flexible tradeoffs between completeness and overhead.

Moreover, the engine
provides a query interface to allow the user to issue queries on the provenance generated by PROV-IO$^+$. Moreover, it includes a visualization module to visualize the provenance (sub)graphs requested by the user. Note that both the query and the visualization need to follow the PROV-IO$^+$ model, which enforces a uniform way to represent the rich provenance information.

{Note that in the preliminary version of the prototype~\cite{PROV-IO}, the user engine only provides a basic query interface to users. As a result,  users have to issue query preimitives one by one to achieve a complicated provenance query. In the current prototype,  PROV-IO$^+$ is further equipped with a set of high-level integrated query APIs for answering typical provenance needs, which can simplify the query complexity and improve the usability  for end users further. For example, in the DASSA use case, to track the backward lineage of an output file, an user only needs to provide the name of the file and the level of predecessor through the integrated query APIs, which will retrieve the provenance information conveniently}. 



\subsubsection{\textbf{Mode \#2: Support for Containerized Workflows}}
\label{sec:containerizer}

To provide provenance support for containerized workflows, PROV-IO$^+$ includes an additional component called \textit{containerizer} besides the components discussed above.

The containerizer engine provides two main functionalities. First, it assembles the target workflows (including their dependencies) as well as the  PROV-IO$^+$ common modules (\S\ref{sec:common-components}) into  container images to be executed on  container platforms.
For example, on an HPC system using Singularity/Apptainer~\cite{singularity}, the containerizer engine first creates the Docker image for the workflow and then converts the Docker image into a Singularity/Apptainer image for execution on compute nodes.

Second, the containerizer engine establishes the  mapping between the directory namespace within the container and the namespace outside the container on the HPC storage nodes, and re-directs the relevant provenance I/O activities to the provenance store for persistency, as shown in the ``Directory Mapping Layer'' and the purple dash lines in Figure~\ref{fig:arch_containerized}). 
In this way, PROV-IO$^+$ can support containerized workflows on HPC systems automatically with little additional efforts. More  implementation details will be discussed in the next section.



\section{PROV-IO$^+$ Implementation}
\label{sec:implementation}
In this section, we discuss  additional implementation details of the major components in the PROV-IO$^+$ framework.


\smallskip
\noindent
{\bf Provenance Tracking.}
To support HDF5 I/O, 
we implement the PROV-IO$^+$ Lib Connector in C and integrate it with the native HDF5 VOL-provenance connector, which follows  a homomorphic design in which each HDF5 native I/O API has a counterpart API~\cite{vol-provenance}. Upon each invocation of an HDF5 native API, 
the counterpart API adds the corresponding virtual data object to a linked list. PROV-IO$^+$ Lib Connector leverages the linked list with locking support to achieve concurrency control on I/O operations on the same data object.
To collect provenance, the PROV-IO$^+$ Library APIs are invoked. 
We collect {\it Agent} information   at the initialization stage of the native HDF5 VOL-provenance connector. {\it Entity} and {\it Activity} classes are tracked at each homomorphic API during the workflow runtime. 

Similarly, to support POSIX I/O, we use  GOTCHA~\cite{GOTCHA} to build a C wrapper layer for POSIX syscall  and invokes the PROV-IO$^+$ Library internally.
Additionally,  the current PROV-IO$^+$ APIs support invoking the PROV-IO$^+$ Library from workflows written in multiple languages including   Python, C/C++, and Java.

{Moreover, to support large-scale ML/AI workflows, we further instrument PyTorch, one of the most popular machine learning framework, with the PROV-IO$^+$  library.
This enables PROV-IO$^+$ to provide more transparency in supporting ML/AI workflows by capturing specific provenance information needed by this category of workflows (e.g., checkpointing information).} 


\smallskip
\noindent
{\bf Provenance Store.}
The Provenance Store is implemented based on Redland \texttt{librdf}~\cite{Redland} to serve as the durable backend of the PROV-IO$^+$ Library. 
We choose Redland because based on our experiences, many other existing RDF solutions are not directly usable in our HPC environments due to compatibility issues in dependent packages and/or operating system (OS) kernels~\cite{Jena, Neo4j, Blazegraph, rya, AnzoGraph}.

We utilize  Redland's  in-memory graph representation and its support for serializing in-memory graph to multiple on-disk RDF formats (e.g., Turtle~\cite{Turtle}, ntriples~\cite{ntriples}, etc.). Redland \texttt{librdf} also supports the integration of multiple databases as the storage backend (e.g., BerkeleyDB, MySQL, SQLite). 
In the current prototype, we store provenance information in the Turtle format directly for simplicity.

To avoid potential data races when serializing from multiple processes to the Provenance Store, PROV-IO$^+$ maintains an in-memory sub-graph for each process and lets the process serialize its own sub-graph to a unique RDF file on disk. The sub-graph files are then parsed and merged into a complete provenance graph.
Since every node in the graph has a globally unique ID (GUID), merging the sub-graphs does not cause unnecessary duplication. Note that this strategy also help performance because no extra inter-process communication or synchronization is needed during workflow execution, and the merging can be performed after workflow execution.

\smallskip
\noindent
{\bf PROV-IO$^+$ User Engine}. 
The user engine supports querying RDF triples with SPARQL, which is a semantic query language to retrieve and manipulate data stored in RDF~\cite{SPARQL}. We use Python scripts as the SPARQL endpoint. Note that depending on different use case scenarios, the query can vary a lot, as will be demonstrated in Section~\S\ref{sec:query_performance}. 
{Note that PROV-IO$^+$ provides highly integrated query APIs for scientists to conveniently retrieve provenance based on their needs. For instance, in DASSA workflow, to query the backward data lineage, the scientists only need to specify the output data object and the level of its predecessor (i.e., how many steps back) to the query API, and the user engine will return the target information if applicable. Similarly, highly integrated and customized query APIs are developed for remaining workflows based on their provenance use cases.}
In the current prototype, 
we utilize  Graphviz~\cite{Graphviz} for RDF graph visualization.

\smallskip
\noindent
{\bf Containerizer Engine.}
The Containerizer Engine is implemented as a set of scripts to enable the provenance support in the containerized environment conveniently. 
For example, 
to support containrized Megatron-LM workflow on the Singularity/Apptainer platform, 
 the Containerizer Engine first creates a Docker image by using the NGC's PyTorch 21.07 as the parent image. 
Besides the Megatron-LM workflow itself, the image also contains the PROV-IO$^+$ library and related dependencies. After the Docker image is created, it is further converted  to a Singularity/Apptainer image in order to run it in the containerized HPC environment.
Moreover, the directory namespace in the container image is mapped to the PROV-IO$^+$ provenance store on the storage nodes for for data persistence. 

Also, since  Singularity/Apptainer provides three running modes (i.e., "run", "exec" and "shell") for different execution scenarios (e.g., interactive jobs and batch jobs),  the Containerizer Engine includes different sets of scripts to support different modes. 
For example, to support running  containerized workflows in the batch mode with the  IBM Spectrum LSF~\cite{lsf} job scheduler, the Containerizer Engine includes scripts to ensure that the configuration parameters of the PROV-IO$^+$ supported containers are consistent with the LSF batch scripts.

\section{Evaluation}
\label{sec:evalution}

{
In this section, we evaluate the prototype of the PROV-IO$^+$ framework in representative HPC environments.
}

First of all, we  introduce the experimental methodology and HPC platforms for non-containerized and containerized workflows, respectively (\S\ref{sec:setup}). 
%
Next, 
we  evaluate   PROV-IO$^+$ with three  non-containerized workflows  (i.e., Top Reco, DASSA and H5Bench) from two perspectives including the tracking performance (\S\ref{sec:tracking_performance}) and the storage requirement (\S\ref{sec:storage_performance}).
Similarly,  we evaluate  PROV-IO$^+$ with one containerized workflow (i.e., Megatron-LM) and measure both the tracking performance  and the storage requirement (\S\ref{sec:megatron-ssc}). 
Moreover, we analyze the impact of containerization on provenance tracking by comparing the tracking overhead in two versions (i.e., with and without containerization) of Megatron-LM (\S\ref{sec:container_impact}). 

\begin{table}[htbp]
    \centering
    \caption{Major Experimental Platforms}
     \vspace{-3pt}
    \begin{tabular}{m{2.4cm} | m{2cm} m{3.2cm}}
        \hline
         & {\bf Cori} & {\bf SAIT SuperCom } \\
        \hline
         Processor & Intel Xeon Phi & AMD EPYC \\ 
         Cores & 622,336 & 204,160 \\
         OS  & Cray Linux & Redhat 8 \\ 
         PFS & Lustre & Lustre \\ 
         Scheduler & Slurm & LSF \\
         Container Runtime & -- & Singularity/Apptainer \\
        \hline
    \end{tabular}
    \label{tab:platforms}
   \vspace{-6pt}
\end{table}

In addition, we  compare PROV-IO$^+$ with a state-of-the-art provenance product (i.e., ProvLake~\cite{ProvLake})  (\S\ref{sec:comparison}), and evaluate the query effectiveness of PRVO-IO for all workflows from the end user's perspective (\S\ref{sec:query_performance}). 

Overall, our experimental results shows that PROV-IO$^+$ can support both non-containerized and containerized workflows effectively. Its tracking overhead is less than 3.5\% in more than 95\% of our experiments,  and it outperforms ProvLake in terms of both tracking and storage overhead. 



\begin{table*}[htbp]
    \centering
    \caption{The provenance needs and the information tracked by PROV-IO$^+$ for three workflows.}
    \vspace{-0.1in}
    \begin{tabular}{m{1.8cm}  m{4.1cm} m{6cm} c c c}
        \hline
        {\bf Workflow} & {\bf Provenance Need} &  {\bf Information Tracked } & {\bf Komadu?} & {\bf ProvLake?} & {\bf PROV-IO$^+$?}\\
        \hline
        Top Reco (Python) & metadata ver. control \& mapping & hyperparameter, preselection, training accuracy & No & Yes & Yes \\
        \hline
        &  file lineage & program, I/O API, file\\
        DASSA& dataset lineage & program, I/O API, dataset  & No & No & Yes \\
        (C++) & attribute lineage & program, I/O API, attr \\
        \hline
        &scenario-1 & I/O API\\
        H5bench & scenario-2 & I/O API, duration & No & No & Yes \\
        (C) &scenario-3 & user, thread, program, file \\
        \hline
        Megatron-LM (Python) & ckpt-config consistency & checkpoint info, loss, model configuration & No & Yes & Yes \\      \hline
    \end{tabular}
    \label{tab:tracking_selection}
    \vspace{-12pt}
\end{table*}

\vspace{-7pt}
\subsection{Experimental Methodology 
}
\label{sec:setup}

\noindent
\textbf{Non-Containerized  Workflows.}
We have evaluated the PROV-IO$^+$ framework for  non-containerized workflows on a state-of-the-art supercomputer named Cori, which is a Cray XC40 supercomputer  deployed at the National Energy Research Scientific Computing Center (NERSC) with a peak performance of about 30 petaflops. 
As shown in Table~\ref{tab:platforms}, 
Cori uses the Slurm job scheduler and do not use container runtime by default.
We conduct experiments on Cori using 64 Intel Xeon ``Haswell'' processor nodes and up to 4096 cores,
 unless otherwise specified.
 The storage backend is a Lustre parallel file system (PFS) with stripe count of 128  and stripe size of 16MB. 


\begin{figure*}[htbp]
    \centering
     \subfloat[Top Reco]{\includegraphics[width=2.3in]{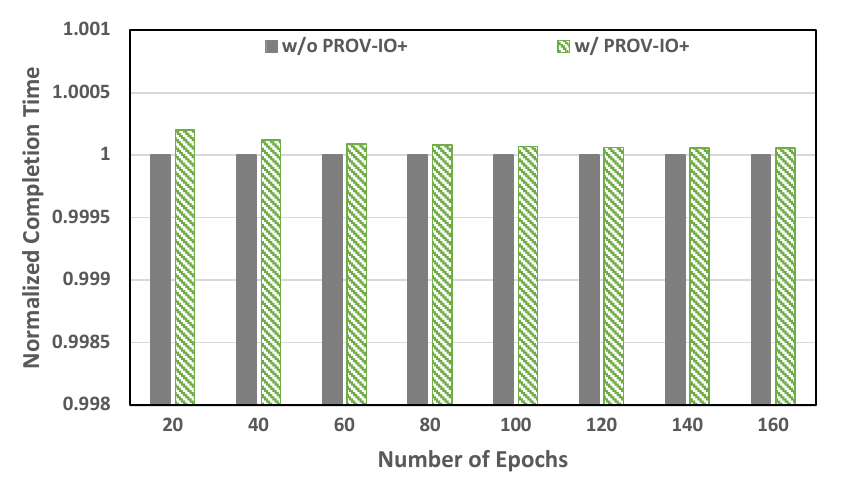}}
    \label{fig:trackinga}
     \subfloat[DASSA]{\includegraphics[width=2.3in]{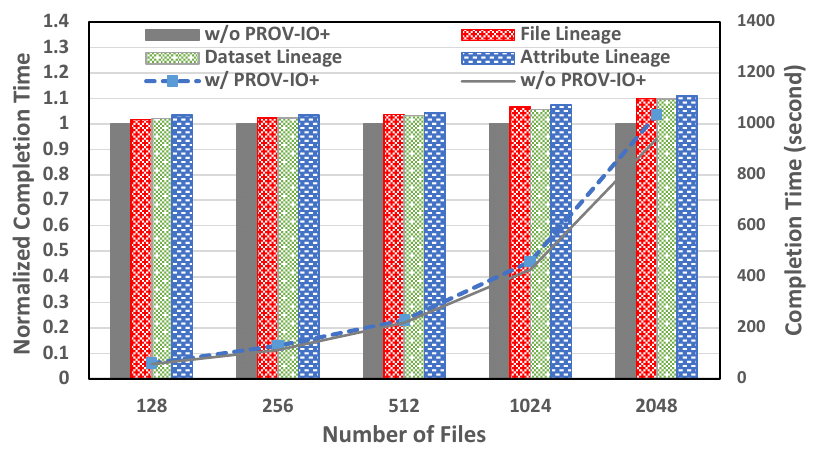}}
             \label{fig:trackingb}
             
     \subfloat[H5bench: write+read]{\includegraphics[width=2.3in]{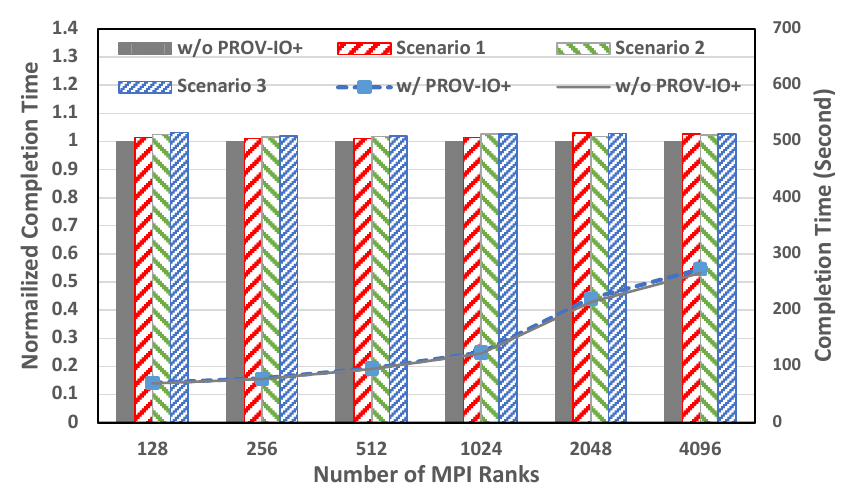}}
             \label{fig:trackingc}
     \subfloat[H5bench: write+overwrite+read]{\includegraphics[width=2.3in]{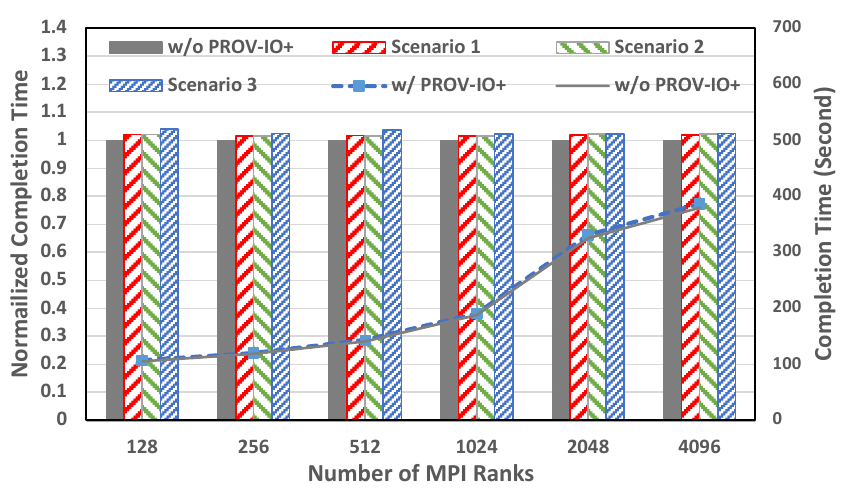}}
        \label{fig:trackingd}
     \subfloat[H5bench: write+append+read]{\includegraphics[width=2.3in]{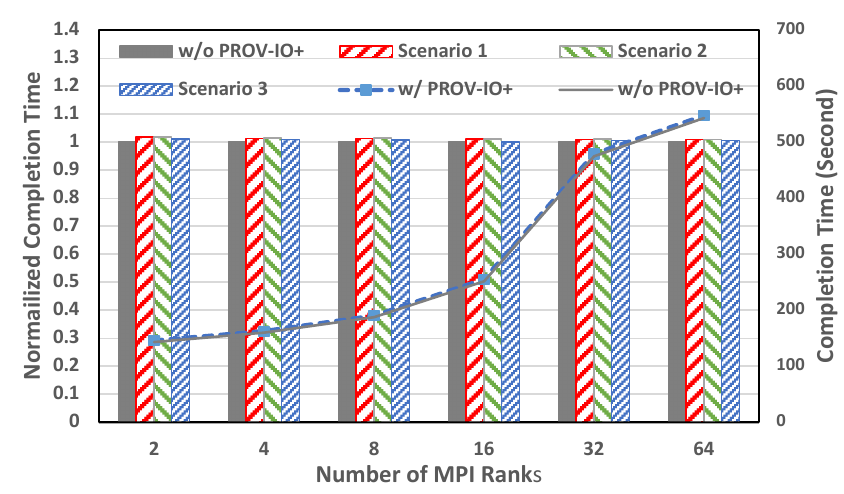}}
        \label{fig:trackinge}

\caption{{\bf Performance of Provenance Tracking.} (a) Top Reco. (b) DASSA (with File, Dataset, Attribute Lineages tracked). (c - e) H5bench-based workflow under  three I/O patterns (i.e., write+read, write+overwrite+read, write+append+read)} 
\label{fig:tracking_overhead}
\end{figure*}

We apply  PROV-IO$^+$ to three representative non-containerized workflows including Top Reco~\cite{TopReco}, DASSA~\cite{DASSA-paper}, and an H5bench-based workflow~\cite{h5bench}. 
As mentioned in \S\ref{sec:casestudies}, the three use cases exhibit diverse characteristics  (e.g.,  various file formats, I/O interfaces, metadata) and provenance needs (e.g., file/dataset/attribute lineage, I/O statistics, metadata versioning). 
We summarize the information tracked by PROV-IO$^+$  in the experiments to meet the provenance needs in
Table~\ref{tab:tracking_selection} and elaborate them in detail in the following subsections.


\smallskip
\noindent
\textbf{Containerized Workflow.}
Besides experimenting with the classic workflows,
we have evaluated the PROV-IO$^+$ framework for one containerized workflow on a supercomputer deployed at Samsung Advanced Institute of Technology (SAIT),  which is an HPE Apollo 6500 Gen10 Plus System. For simplicity, we call the system  SuperCom in the rest of the paper. Similar to  Cori, SuperCom uses Lustre as the parallel file system. 
Different from Cori, SuperCom's job management is based on a combination of IBM LSF scheduler and Singularity/Apptainer, which is a container runtime designed for HPC environments~\cite{singularity}.
A  detailed comparison between the two platforms (i.e., Cori and SuperCom ) is summarized in Table~\ref{tab:platforms}. 

We apply PROV-IO$^+$ to the representative deep learning  workflow Megatron-LM~\cite{Megatron1}, which exhibits unique characteristics and provenance needs as discussed in \S\ref{sec:megatron-lm} and summarized in Table~\ref{tab:tracking_selection}. We containerize the workflow through  the PROV-IO$^+$ containerizer engine and leverage eight NVIDIA A100 GPUs on SAIT SuperCom to accelerate the training. 
For clarity, we present the  evaluation results on SAIT SuperCom in \S\ref{sec:megatron-ssc}. 

\smallskip
\noindent
\textbf{Containerization Impact Analysis.}
In addition to experimenting on Cori and SuperCom, we have used Google Cloud Platform (GCP) to study the impact of containerization on provenance tracking. 
We use GCP because  Cori and SuperCom  are customized for supporting LBNL's and Samsung's missions respectively, and we cannot modify their runtime environment (e.g., adding or removing Singularity/Apptainer) conveniently. By leveraging GCP, we can build the necessary system environments for running both non-containerized and containerized workflows and conduct a fair comparison on the same infrastructure.

More specifically, we use the GCP Deep Learning virtual machines (VMs) with 32 vCPUs, 120 GB DRAM, and 4 NVIDIA T4 GPUs for the comparison experiments. And we apply PROV-IO$^+$ in two different modes for provenance tracking on  two versions of Megatron-LM (i.e., non-containerized  and containerized) respectively. We discuss the comparison results in \S\ref{sec:container_impact}.
 

\subsection{Performance of Provenance Tracking}
\label{sec:tracking_performance}

In case of {\bf Top Reco}, the scientists need the mapping between configurations  and the  training performance. 
Therefore, PROV-IO$^+$ tracks three domain-specific items (e.g., model  hyperparameters,  dataset preselections, and training accuracy) 
based on the extensible class defined in the PROV-IO$^+$ model. 
To track the mapping between workflow configuration and training accuracy, we instrument the workflow's training loop with PROV-IO$^+$ APIs and  record the training accuracy at the end of each epoch, and add the training accuracy to the provenance graph as a property of
configurations. 
In addition, we vary the number of training epochs to see how the  performance scales.
Note that Top Reco is a single process workflow.


 Figure~\ref{fig:tracking_overhead}(a) shows the  performance for Top Reco. The y-axis is the normalized completion time (starting with 0.998),  while the x-axis is the number of training epoch (roughly equivalent to training time). 
The grey bars are the baseline without provenance, and the green bars show the performance with PROV-IO$^+$ enabled. 
 We can see that the tracking overhead is negligible overall with a maximum of 0.02\%. 
 The overhead with a shorter training time is relatively high, which is mostly caused by the  latency of Redland. As the number of training epoch increases, the overhead of PROV-IO$^+$ decreases almost linearly because PROV-IO$^+$ tracks a constant amount of information. 

In case of {\bf DASSA},  the scientists need the backward lineage of data products in different granularity. As shown in the second column of Table~\ref{tab:tracking_selection}, PROV-IO$^+$ tracks the information of user, program, file, dataset, or attribute for different lineage needs based on the PROV-IO$^+$ model (\S\ref{sec:model}). We follow a similar configuration as the domain scientists' by using 
32 compute nodes and up to  2048 input files (1.35TB in total).

 Figure~\ref{fig:tracking_overhead}(b) shows the tracking performance for DASSA. The x-axis means the number of input files; the y-axis on the left and right sides show the normalized  completion time and the raw completion time (in second), respectively. The grey bars represent the normalized  baseline without PROV-IO$^+$,
and the red, green and blue bars represent the normalized completion time under three usage scenarios (i.e., ``File Lineage'', ``Dataset Lineage'' and ``Attribute Lineage'') where different provenance  granularity are enabled (e.g.,  for ``File Lineage'' we enable ``program'', ``I/O API'' and ``file'' tracking). 
The solid grey line means the average baseline completion time (in second) without provenance tracking, while the  dashed blue line represents the worst case raw completion time with PROV-IO$^+$ under all  scenarios.  

We can see the max overhead occurred when tracking the attribute lineage of the entire 2048 files,  which is about 11\%. This is because DASSA  heavily relies on HDF5 attributes. To access an attribute, the program first needs to open the file and the dataset containing it, which incurs more I/O operations to track. But overall, PROV-IO$^+$ incurs reasonable overhead in  DASSA (from 1.8\% to 11\%).
This is expected because  DASSA  does not require heavy I/O API  tracking. 
In other words, PROV-IO$^+$ is  efficient for tracking the backward lineage in file, dataset, and attribute granularity.

\begin{figure*}[htbp]
    \centering
     \subfloat[Top Reco]{\includegraphics[width=2.3in]{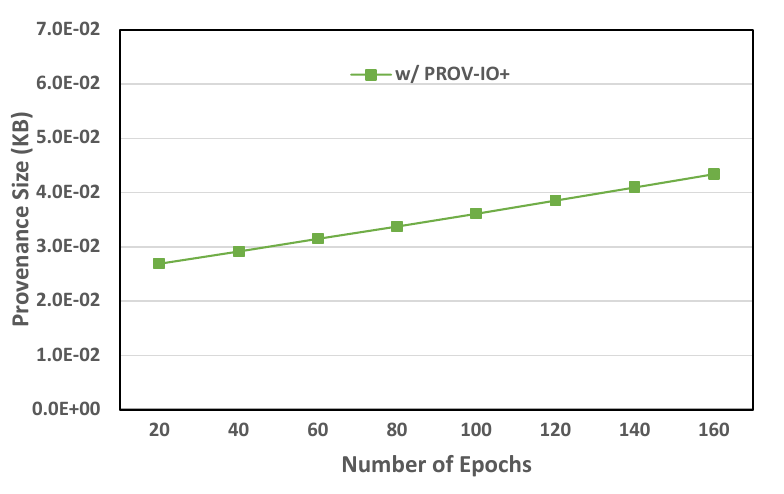}}
    \label{fig:storagea}
     \subfloat[DASSA]{\includegraphics[width=2.3in]{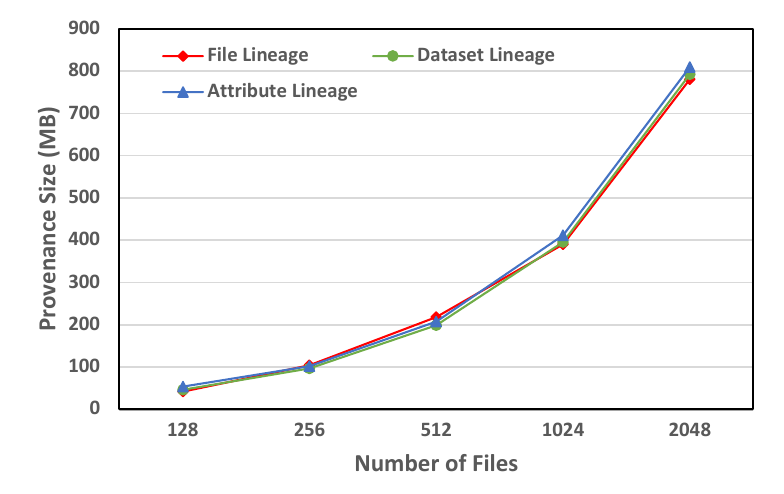}}
             \label{fig:storageb}
             
     \subfloat[H5bench: write+read]{\includegraphics[width=2.3in]{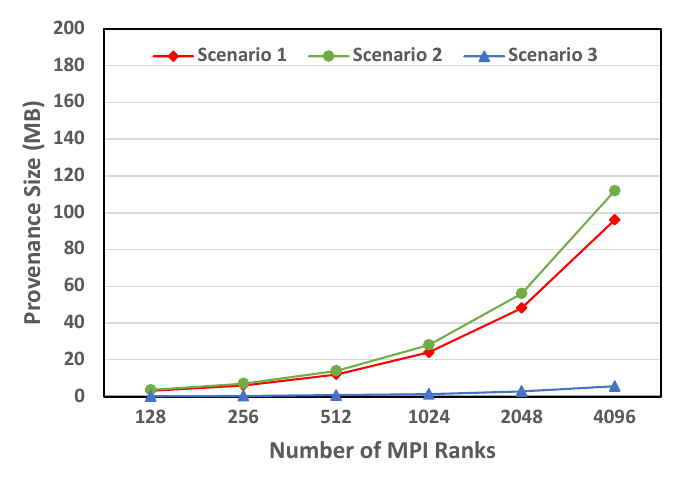}}
             \label{fig:storagec}
     \subfloat[H5bench: write+overwrite+read]{\includegraphics[width=2.3in]{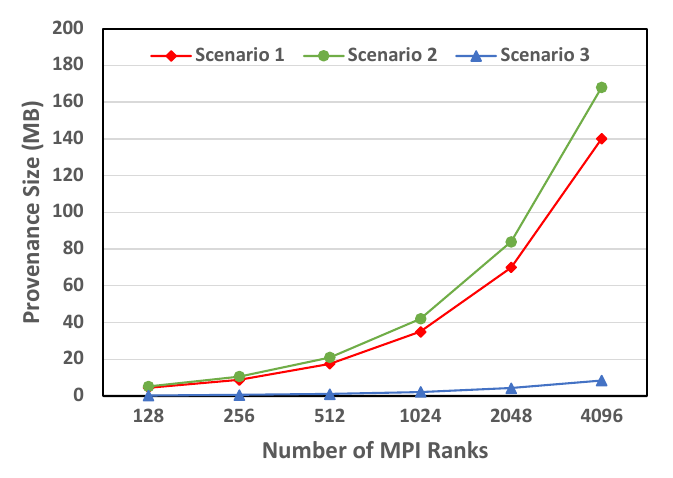}}
        \label{fig:storaged}
     \subfloat[H5bench: write+append+read]{\includegraphics[width=2.3in]{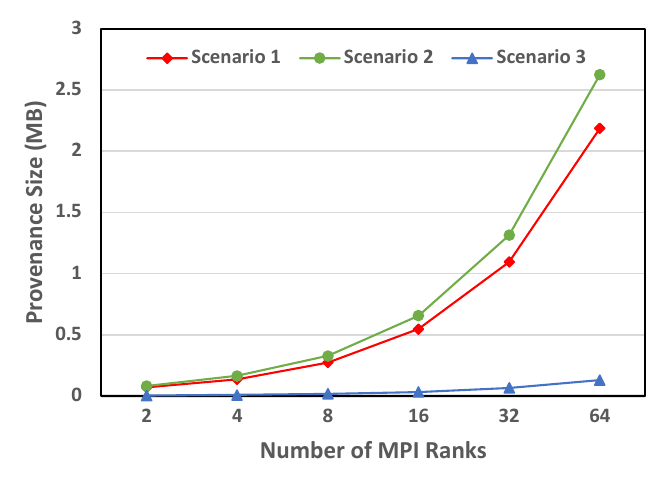}}
        \label{fig:storagee}

\caption{{\bf Storage of Provenance Tracking.} (a) Top Reco. (b) DASSA (with File, Dataset, Attribute Lineages tracked). (c - e) H5bench-based workflow under  three I/O patterns (i.e., write+read, write+overwrite+read, write+append+read)} 
\label{fig:storage_overhead}
\end{figure*}

In the {\bf H5bench} based workflow, the scientists need the data usage and I/O statistics in general. We consider three different usage scenarios with different needs. As summarized in Table~\ref{tab:tracking_selection}, {\it scenario-1}  tracks the total number of I/O APIs;  {\it scenario-2} tracks both the I/O API count and their duration for bottleneck analysis; {\it scenario-3}  tracks the users and threads that modify the file. Moreover, for each scenario, we consider three different I/O patterns including: {\it write-read}, {\it write-overwrite-read}, and {\it write-append-read}. In (c) and (d), we run the workflow with  128 to 4096 MPI processes. In (e), since the append operations from a large amount of MPI processes can easily overwhelm the memory buffer for appending and lead to out-of-memory (OOM) errors, we reduce the number of MPI processes ( 2 to 64). 
Also, based on the observation that the computation time of many HPC applications may vary from dozens to thousands of seconds per I/O operation, we introduce a relatively modest computation time of 25 seconds per step in the experiments.

Figure~\ref{fig:tracking_overhead} (c) (d) (e) show the tracking performance under three different I/O patterns (i.e., ``write+read'', ``write+overwrite+read'', ``write+append+read'').
The x-axis stands for the number of MPI ranks. The left y-axis is the normalized completion time and the right y-axis is the raw completion time in second. The grey bars represent the baseline while the three types of colored bars stand for the performance  of different provenance usage scenarios mentioned in Table~\ref{tab:tracking_selection}(red  for ``scenario 1'', green  for ``scenario 2'', blue  for ``scenario 3''). The grey solid line is the average baseline completion time, while the blue dash line is the worst-case raw completion time with PROV-IO$^+$ enabled.

Overall, we find that PROV-IO$^+$ incurs reasonable amount of overhead (i.e., ranging from 0.5\% to 4\%) even under heavy I/O operations (3.9TB data with 4096 MPI ranks). 
In particular, the PROV-IO$^+$ overhead under the ``write-append-read'' I/O pattern (Figure~\ref{fig:tracking_overhead} (c)) is  minimal (around 0.5\%).
This is because the HDF5 I/O operation under this pattern  takes  more computation time than under the other two patterns to determine the append offset and memory range, which makes the PROV-IO$^+$ overhead more negligible.
Also, by comparing scenario-1 and scenario-2, we find that tracking the I/O API duration  introduce little additional overhead. This is reasonable because the timing information 
can be piggybacked with the I/O API tracking which  dominates the overall tracking time.

\vspace{-9pt}
\subsection{Storage Requirements}
\label{sec:storage_performance}

The storage requirement of PROV-IO$^+$ is directly related to the amount and the class of information tracked.
Specifically, 
the storage overhead may increase in two ways: (1) the size of a single provenance record may increase (e.g., adding timing information will increase size of an I/O API record); (2) the total number of records in a provenance file may increase (e.g., tracking thread information will create a number of thread records).
We summarize the storage performance of PROV-IO$^+$ for the three workflows in Figure~\ref{fig:storage_overhead}. 

Figure~\ref{fig:storage_overhead}(a) shows the Top Reco case. 
The x-axis represents the number of epochs and the y-axis is the provenance size (KB). We can see that the provenance size is negligible. This is because PROV-IO$^+$ allows users to specify the target provenance precisely without incurring unnecessary overhead. 
It also scales linearly since the number of new nodes added to provenance graph is the same as the increment in training epochs.

Figure~\ref{fig:storage_overhead}(b) shows the DASSA case. 
The x-axis is the number of input files while the y-axis represents the provenance size (MB).  Lines in three different colors represent File Lineage, Dataset Lineage and Attribute Lineage, respectively. We can see that the storage requirement  varies from 40 MBs (with 128 input files) to about 800 MBs (with 2048 files) with linear scalability (note that the x-axis increases by a multiple of 2). Although DASSA heavily relies on attributes, the storage overhead in the three usage scenarios is similar. This is because I/O API is still the dominant part in all scenarios. 
Even though the number of file and dataset is far less than attribute in DASSA input data, when compared to number of APIs involved in the workflow, their contribution to storage overhead is insignificant.


Figure~\ref{fig:storage_overhead} (c)(d)(e) shows the H5bench-based workflow
with three different I/O patterns.
The x-axis represents number of MPI ranks and the y-axis stands for provenance size in MBs. Note that x-axis also increases by a multiple of 2. Lines in three different colors represents  three different provenance usage scenarios (Table~\ref{tab:tracking_selection}). We can see the provenance size varies from a few KBs to 168 MBs. Among the three I/O patterns, ``write+overwrite+read'' has the highest storage overhead under usage scenario 2. This is because the pattern includes one more I/O application (i.e., overwrite) than ``write+read'' and has much more MPI processes contributing to  provenance graph than  ``write+append+read''.
Moreover, scenario 2 also has the largest amount of tracked information (I/O API and their duration). 
Note that the storage overhead in this workflow also scales linearly.

In summary, because of the flexibility of the fine-grained PROV-IO$^+$ model, 
PROV-IO$^+$'s storage overhead is reasonable for all the use cases evaluated.
\subsection{Experiments with Containerized  Workflow}
\label{sec:megatron-ssc}


In this section, we introduce our experiments with a containerized workflow (i.e., Megatron-LM~\cite{Megatron1}) on the Samsung supercomputer (SuperCom). 

  Megatron-LM   supports multiple pretraining models (\S\ref{sec:megatron-lm}), and we configure Megatron-LM to use GPT-2 as the pretraining model in this set of experiments based on the need of the domain scientists.  The training dataset is WikiText103~\cite{WikiText103} which is provided by the  Megatron-LM authors~\cite{Megatron1}.  We enable both model parallelism and data parallelism (\S\ref{sec:megatron-lm}) in the workflow for  experiments.

As mentioned in \S\ref{sec:megatron-lm}, the major provenance need in the Megatron-LM  use case is to ensure the consistency between the checkpoint and the workflow configurations. Therefore, we track detailed checkpoint information pertaining to a pretraining process (e.g.,  the  path of the checkpoint file) as well as a variety of relevant configuration parameters (e.g., the number of GPUs, the batch size). The configuration information is tracked once at the beginning of workflow execution as the information remains invariant throughout the workflow execution, while the checkpoint information is tracked transparently by instrumenting PyTorch at the end of the workflow execution. In addition, we record the GPT-2 training loss    at the end of each training iteration. We change the number of training iterations in the experiments to measure how the tracking performance and storage requirement scales. 
We report the measurement results as follows.

\begin{figure}[htbp]
\centering
\subfloat[Tracking Performance.]{\includegraphics[width=1.75in]{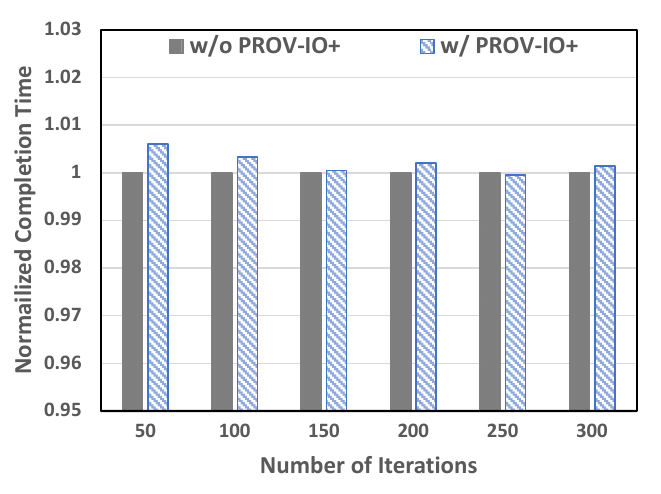}%
\label{fig:megatron-lm-tracking}}
\hfil
\subfloat[Storage Requirement.]{\includegraphics[width=1.75in]{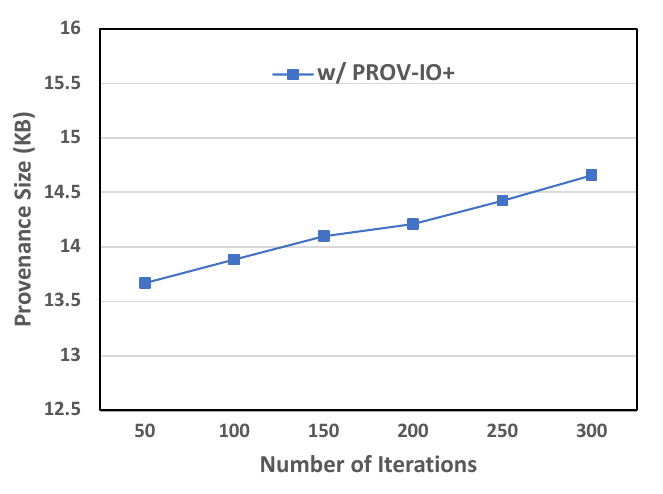}%
\label{fig:megatron-lm-storage}}
\vspace{-4pt}
\caption{{\bf Performance of PROV-IO$^+$ on Megatron-LM with checkpoint path, training loss and configuration tracked.} }
\label{megatron-lm-result}
\end{figure}

Figure~\ref{fig:megatron-lm-tracking} shows the provenance tracking performance. The y-axis is the normalized workflow completion time, and the x-axis is the number of training iteration. We use the grey bar to represent the baseline without provenance tracking, and the blue bar stands for workflow completion time with provenance enabled. For each experiment, we repeat it five times and calculate the average performance value. The result shows that the maximum tracking overhead is about 0.6\% when the training iteration is set to 50. When the  number of iteration increases, PROV-IO$^+$'s tracking overhead tends to be negligible, which implies that PROV-IO$^+$ is scalable in terms of tracking performance. 

Figure~\ref{fig:megatron-lm-storage} shows the storage requirement of tracking Megatron-LM with PROV-IO$^+$. The y-axis is the provenance size (KB), and the x-axis is the number of training iteration. The result shows that, to track the checkpoint-configuration consistency information, the provenance size is negligible in general (e.g., less than 15 KB in all experiments). The size of the provenance information scales almost linearly as the number of  training iterations increases, mainly because the {training loss} is recorded at the end of each training iteration.

In summary, to track the  necessary provenance  for maintaining the checkpoint-configuration consistency in  containerized Megatron-LM, PROV-IO$^+$ introduces small tracking  and negligible storage consumption.




\subsection{Impact of Containerziation on Provenance Tracking}
\label{sec:container_impact}

In this section, we analyze the impact of containerization on PROV-IO$^+$'s provenance collection. As mentioned in \S\ref{sec:setup}, we leverage the GCP platform for the comparison experiments because we can setup different runtime environments for both containerized and non-containarized workflows on GCP.
We  apply PROV-IO$^+$ in two different modes for tracking two versions of Megatron-LM (i.e., non-containerized  and containerized) on GCP respectively.


To validate the impact of containerization, we  execute the non-containerized version of Megatron-LM workflow and the containerized version separately on different GCP VMs. This is to ensure that there is no interference between the executions of the two versions. The provenance information tracked is the same as described in \S\ref{sec:megatron-ssc}. We reduce the scale of the workflow to meet the VM's resource constraints (e.g., vCPUs and memory).

\begin{figure}[htbp]
\centering
\subfloat[Non-containerized workflow]{\includegraphics[width=1.75in]{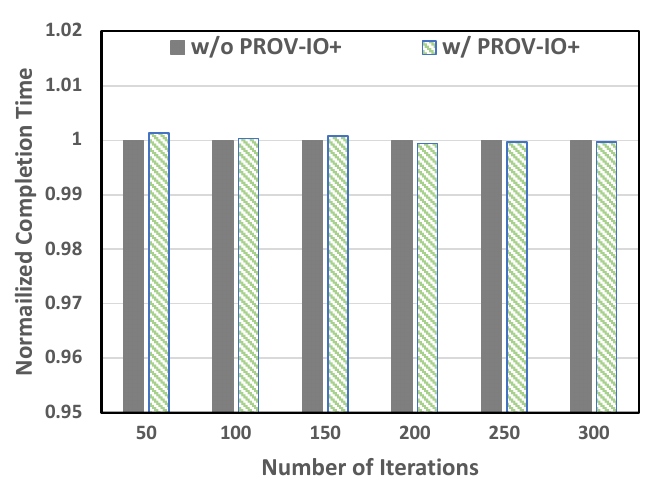}%
\label{fig:non-containerized}}
\subfloat[Containerized workflow]{\includegraphics[width=1.75in]{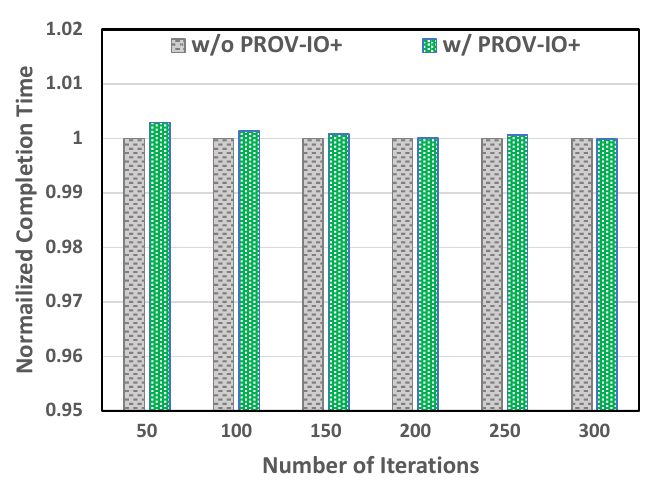}%
\label{fig:containerized}}
\vspace{-4pt}
\caption{{\bf A Comparison of PROV-IO$^+$ tracking overhead in non-containerized and containerized Megatron-LM.} }
\label{fig:container_impact}
\end{figure}

The performance of PROV-IO$^+$ on non-containerized Megatron-LM  and containerized Megatron-LM are shown in Figure~\ref{fig:non-containerized}  and Figure~\ref{fig:containerized}, respectively. In both cases, the y-axis
is the normalized workflow completion time, and the x-
axis is the number of training iteration. 
By comparing Figure~\ref{fig:non-containerized} and Figure~\ref{fig:containerized}, we can see that in both cases  PROV-IO$^+$ incurs little overhead, especially when the number of iterations is large. 
This suggests that containerization has little impact on PROV-IO$^+$, and both modes of PROV-IO$^+$ can support provenance tracking efficiently.



In conclusion, our experiments on three different platforms (i.e., Cori in \S\ref{sec:tracking_performance}, SuperCom 
 in \S\ref{sec:megatron-ssc}, and GCP \S\ref{sec:container_impact}) shows that PROV-IO$^+$'s provenance tracking performance has little dependence on the platforms,  and the  overhead is consistently low across different execution platforms.  




\vspace{-16pt}

\begin{figure*}[htbp]
    \centering
     \subfloat[Tracking Overhead - 20 Config Fields.]{\includegraphics[width=2.3in]{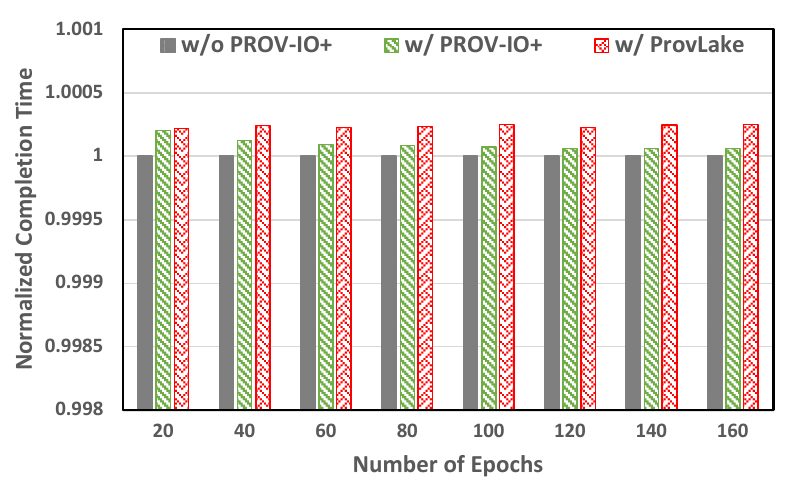}}
     \subfloat[Tracking Overhead - 40 Config Fields.]{\includegraphics[width=2.3in]{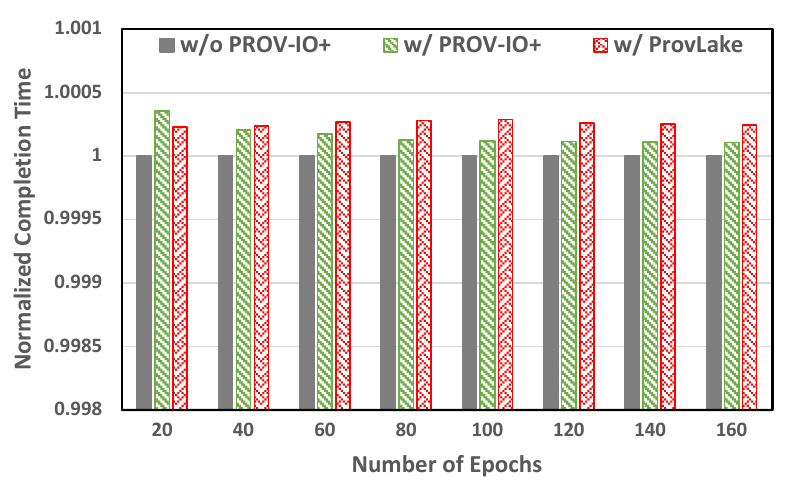}}
     \subfloat[Tracking Overhead - 80 Config Fields.]{\includegraphics[width=2.3in]{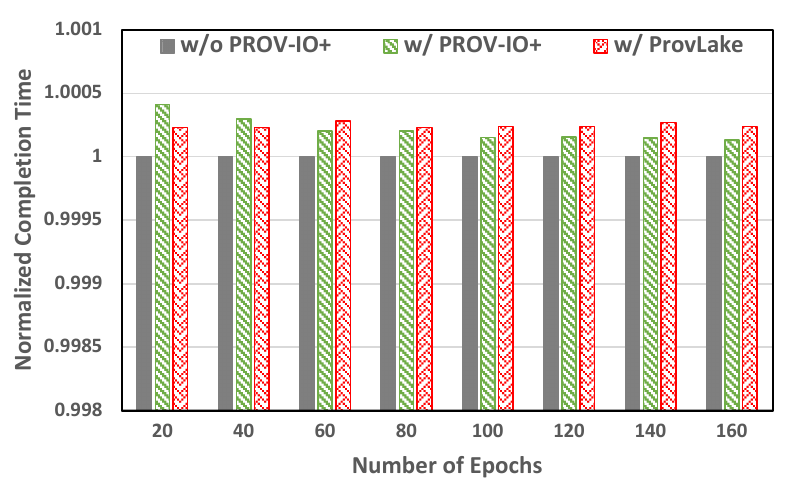}}
             \label{fig:provlakec}
             
     \subfloat[Storage Overhead - 20 Config Fields.]{\includegraphics[width=2.3in]{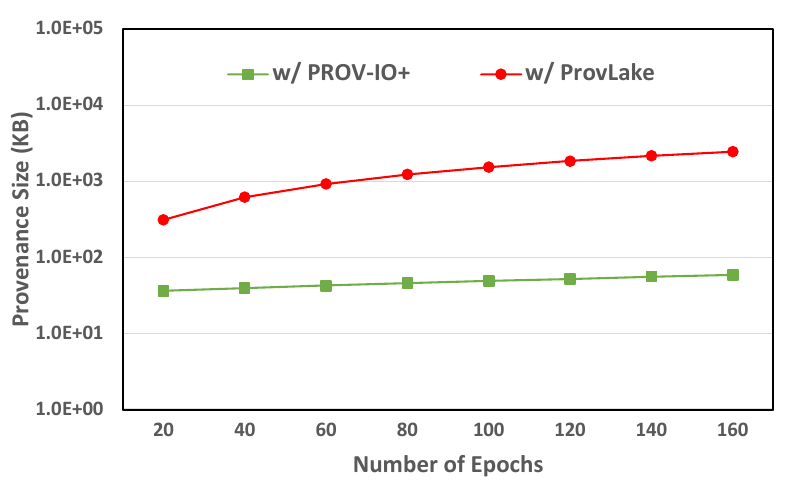}}
        \label{fig:provlaked}
     \subfloat[Storage Overhead - 40 Config Fields.]{\includegraphics[width=2.3in]{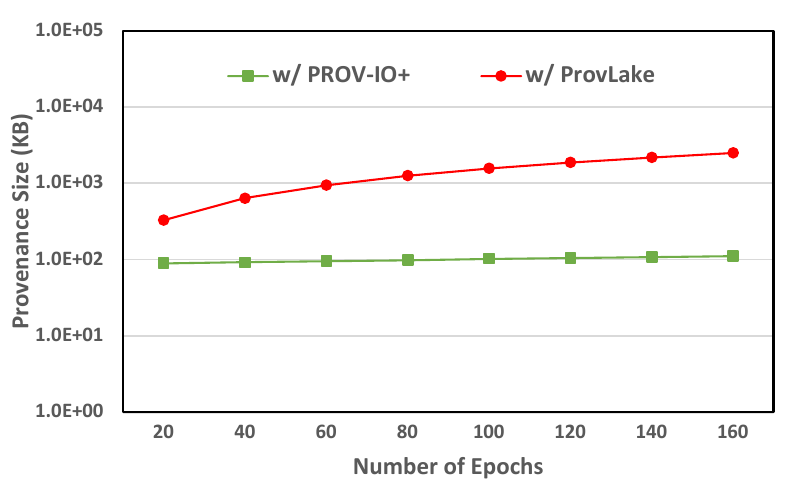}}
        \label{fig:provlakee}
    \subfloat[Storage Overhead - 80 Config Fields.]{\includegraphics[width=2.3in]{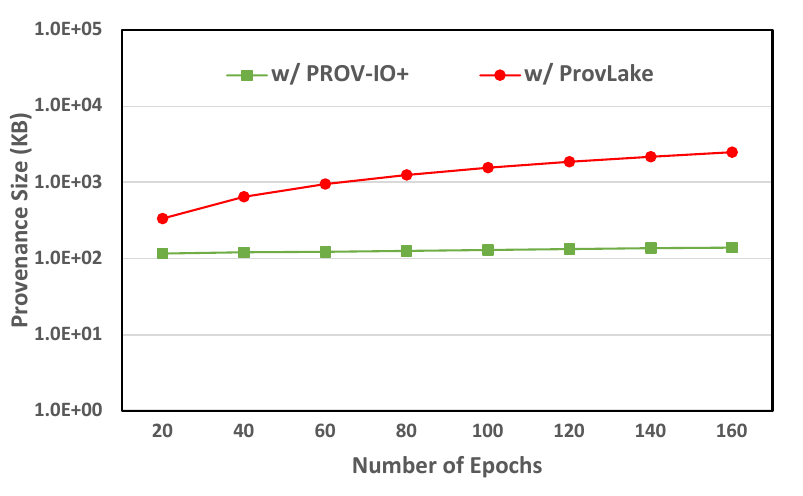}}
         \label{fig:provlakef}

\caption{{\bf A performance comparison between PROV-IO$^+$ and ProvLake.} } 
\label{fig:provlake_comparison}
\end{figure*}

\vspace{-0.2in}
\subsection{Comparison with Other  Frameworks}
\label{sec:comparison}


In this section, we compare PROV-IO$^+$ to state-of-the-art provenance systems.
Table~\ref{tab:comparison} shows the basic characteristics of Komadu~\cite{Komadu},
ProvLake~\cite{ProvLake}, and PROV-IO$^+$. 
We can see that all three frameworks are derived from  the base PROV-DM  model, which makes the comparison fair. 
On the other hand,  Komadu only supports Java programs and 
ProvLake only supports Python, which makes them incompatible with many C/C++ based scientific workflows (e.g., DASSA and H5bench). 
Note that PROV-IO$^+$'s C/C++ interface is designed for integration with major HPC I/O libraries. Once the I/O library is integrated with PROV-IO$^+$ (e.g., HDF5), the provenance support is mostly transparent to the workflow users, i.e., users can control the rich provenance features through a configuration file without manually modifying their source code with APIs. 
Neither Komadu nor ProvLake support such capability or transparency.


\begin{table}[htbp]
    \centering
    \caption{Basic Characteristics of Three  Frameworks.}
        \vspace{-0.1in}
    \begin{tabular}{m{1.5cm} m{1.5cm} m{1.5cm} m{2.5 cm}}
        \hline
         &  {\bf Komadu} &  {\bf ProvLake } & {\bf PROV-IO$^+$} \\
        \hline
        Base model & PROV-DM  & PROV-DM & PROV-DM \\
        Language & Java & Python & {\small C/C++,Python,Java} \\
        Transparency & No & No & Hybrid \\
        \hline
    \end{tabular}
    \label{tab:comparison}
    \vspace{-6pt}
\end{table}

Since ProvLake 
has outperformed Komadu based on a previous study~\cite{EfficientRuntimeCapture}, we focus on the comparison with ProvLake.
Because ProvLake does not support C/C++ workflows, we cannot apply it to DASSA and H5bench. 
Therefore, we compare the two provenance tools using Python-based Top Reco in the rest of this section.


Different from PROV-IO$^+$  which is I/O-centric, ProvLake is  'process-oriented'.  
Specifically,  ProvLake creates records based on the execution steps of a workflow, and the provenance data are maintained as attribute or property of individual steps. On the contrary, PROV-IO$^+$ is not limited to the execution steps of the workflow. For example, it can track a task in the workflow, an I/O operation invoked by a task,  a data object involved in the I/O operation, etc., all of which are further correlated via the relations defined by the PROV-IO$^+$ model (\S\ref{sec:model}). 
Such flexibility and richness is not available in ProvLake.


To make the comparison with ProvLake fair, we use the same instrument points in the Top Reco workflow for both tools. Specifically, we instrument Top Reco at its GNN training loop and track the training accuracy at the end of each epoch to corresponding provenance records.
Since the workflow configuration is never changed during the entire workflow, we only add it to ProvLake's record once at the beginning of the workflow. 
In addition, to be representative, we track three different numbers of  configurations (i.e., 20, 40, and 80).

\begin{table*}[htbp]
    \centering
    \caption{Example Queries. The diverse provenance needs can be satisfied by  a few simple queries effectively.}
    \vspace{-0.1in}
    \begin{tabular}{c|c|l|c}

        \hline
        {\bf Workflow} & {\bf Provenance Need} & {\bf Query Statement (SPARQL)} & {\bf $\#$ of Statements in Query} \\
        \hline
        & \cellcolor{Gray}& \cellcolor{Gray}1: data\_object\_a prov:wasAttributedTo ?program. & \cellcolor{Gray} 3*N \\
        DASSA & \cellcolor{Gray} file/dataset/attribute lineage & \cellcolor{Gray}2: ?data\_object prov:wasAttributedTo ?program; & \cellcolor{Gray} {\small (where N is backward}\\
        & \cellcolor{Gray}& \cellcolor{Gray}3: $\>\>\>\>\>\>\>\>\>\>\>\>\>\>\>\>\>\>\>\>\>\>\>$ provio:wasReadBy ?IO\_API. & \cellcolor{Gray} {\small  propagation steps)}\\
        \hline
        & scenario-1  & 4: ?IO\_API prov:wasMemberOf prov:Activity; & 1\\
        & \cellcolor{Gray}scenario-2   & \cellcolor{Gray}5: ?IO\_API prov:wasMemberOf prov:Activity; & \cellcolor{Gray}2 \\
        H5bench  &\cellcolor{Gray} & \cellcolor{Gray}6: $\>\>\>\>\>\>\>\>\>\>\>\>\>\>\>$ provio:elapsed ?duration.&\cellcolor{Gray} \\
        & & 7: file\_a prov:wasAttributedTo ?program. & \\
        & scenario-3  & 8: ?program prov:actedOnBehalfOf ?thread. & 3 \\
        & & 9: ?thread prov:actedOnBehalfOf ?user. & \\
        \hline
        Top Reco & \cellcolor{Gray}metadata version control \& mapping & 10: \cellcolor{Gray}?configuration ns1:Version ?version; & \cellcolor{Gray}2\\
        & \cellcolor{Gray}&  \cellcolor{Gray}11: $\>\>\>\>\>\>\>\>\>\>\>\>\>\>\>\>\>\>\>\>\>\>\>\>\>\>$provio:hasAccuracy ?accuracy. & \cellcolor{Gray}\\
        \hline 
        Megatron-LM &  ckpt-config consistency & 12: ?batch\_size ns1:hasValue 256; & 2 \\
        &   &  13: $\>\>\>\>\>\>\>\>\>\>\>\>\>\>\>\>\>\>\>\>\>$ prov:influenced ?checkpoint\_path & \\
        \hline 
    \end{tabular}
    \label{tab:query_example}
   \vspace{-16pt}
\end{table*}

Figure~\ref{fig:provlake_comparison}(a),(b),(c) compares the provenance tracking performance of the two systems where y-axis is normalized completion time.  Figure~\ref{fig:provlake_comparison} (d),(e),(f) shows the storage overhead  where y-axis is size in KB. In all figures x-axis is the number of configurations. In (a)(b)(c), grey bars stand for the baseline without provenance tracking, green bars show the normalized performance with PROV-IO$^+$, and  red bars show the performance with ProvLake. In (d)(e)(f), green lines stand for PROV-IO$^+$ provenance file size and red lines stand for ProvLake provenance file size.

As shown in Figure\ref{fig:provlake_comparison}(a)(b)(c), both frameworks incur negligible tracking overhead (e.g., less than 0.025\%) and the PROV-IO$^+$ overhead is even lower than ProvLake for most cases. 
Similarly, as shown in Figure\ref{fig:provlake_comparison}(d)(e)(f), PROV-IO$^+$ always incurs less storage overhead, regardless of the number of  configuration fields tracked.
This is mainly because ProvLake has to track more irrelevant workflow information not needed in the use case.






\vspace{-0.05in}
\subsection{Query  Effectiveness}
\label{sec:query_performance}
As mentioned in \S\ref{sec:implementation}, PROV-IO$^+$ supports provenance query with visualization.
Table~\ref{tab:query_example} summarizes the  queries used to answer the diverse provenance needs of the three workflow cases. We can see that the provenance  can be queried effectively and efficiently using a few simple SPARQL statements in general. Since the number of queries involved is small, the query time overhead is negligible in our experiments.  We discuss each case in more details   below.


In DASSA, to get the backward lineage of a data product, we can start with  the program which generated the data product and look for its input data. The same procedure can be repeated as needed.  For example, DASSA may convert ``WestSac.tdms'' into ``WestSac.h5'' with program ``tdms2h5'', and then use ``decimate'' to process ``WestSac.h5'' into data product ``decimate.h5''. 
{To get the backward lineage of ``decimate.h5'', in the query, we first retrieve with keywords ``decimate.h5 prov:wasAttributedTo ?program'' to locate program ``decimate''. Next, in the same query, we add statement ``?file wasAttributedTo ?program'' to retrieve that program's input file ``WestSac.h5'' which is the first level predecessor of ``decimate.h5''. We can further expand the query by adding similar statements to locate ``decimate.h5'''s earlier predecessors (e.g., ``WestSac.tdms'').}

\begin{figure}[htbp]
\centering
\centerline{\includegraphics[width=3in]{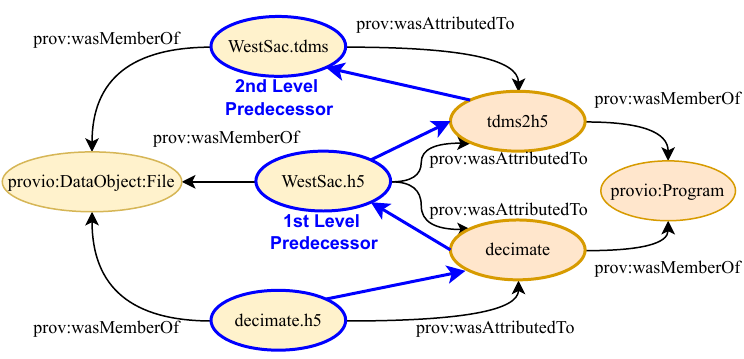}}
    \vspace{-0.1in}
\caption{{\bf An Example of DASSA Data Lineage  by PROV-IO$^+$.} The  graph follows the PROV-IO$^+$  model; the data lineage is highlighted in blue.}
\label{fig:lineage_example}
\end{figure}

As summarized in Table~\ref{tab:query_example}, for each backward step, we only need three query statements. 
Figure~\ref{fig:lineage_example} shows the visualization of this example, which follows the PROV-IO$^+$ provenance model (\S\ref{sec:model}) and highlights the queried data lineage in blue.
Other types of lineages (e.g., dataset and attribute) can be queried and visualized in the same way.

\vspace{-4pt}

Similarly, in H5bench, we have three types of provenance needs (i.e., the scenarios described in \S\ref{sec:tracking_performance}) which can be answered  using 1, 2, 3 SPARQL statements  respectively. In Top Reco, the metadata versioning and mapping information can be queried in 2 statements. 
Note that the provenance needs are diverse across the real use cases, but the number of queries needed is consistently small. This elegant result suggests that   PROVI-IO$^+$  is effective for scientific data on HPC systems.




%

\begin{figure}[htbp]
\centering
\centerline{\includegraphics[width=3in]{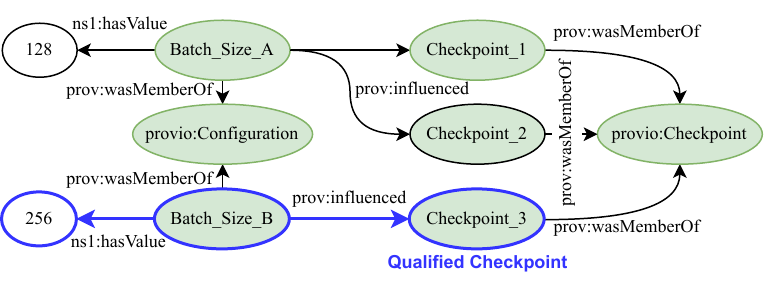}}
    \vspace{-0.1in}
\caption{{\bf An Example of Megatron-LM query  by PROV-IO$^+$.} The  graph follows the PROV-IO$^+$  model; the query path is highlighted in blue.}
\label{fig:megatron_query_example}
\end{figure}

In case of Megatron-LM, users want to identify the checkpoint which is consistent with the configuration for the follow-up training process. Figure~\ref{fig:megatron_query_example} shows a scenario where there are two different batch sizes used during the previous pretraining (i.e., ``Batch\_Size\_A'' is 128, and ``Batch\_Size\_B'' is 256), and there are three checkpoints generated based on the two batch sizes (i.e., ``Checkpoint\_1'', ``Checkpoint\_2'', ``Checkpoint\_3''). Assume the user wants to continue a GPT pretraining process which has a batch size of 256 with one of the existing checkpoints (i.e., ``Checkpoint\_3''), s/he can query the provenance  with  as few as 2 lines of SPARQL statements, as shown in the last row of Table~\ref{tab:query_example}. Moreover, the user can also add advanced conditions to the query to filter out the feasible checkpoint with the best quality (e.g., a checkpoint with certain training loss).

\vspace{-4pt}

\vspace{-10pt}

\vspace{-5pt}
\section{Discussion}
\label{sec:discussion}

The design of the PROV-IO$^+$ tool is driven by the needs of the domain scientists using four scientific workflows. Given the diversity of science, it is likely that the prototype cannot directly address the  unique provenance queries of all scientists. 
We plan to collaborate with more domain scientists to identify additional needs and refine PROV-IO$^+$ accordingly. 
For example,  researchers from INRIA~\cite{inria}  are interested in porting PROV-IO$^+$ on their edge devices with limited hardware resource. Similarly, HPE~\cite{hplab} researchers are interested in integrating PROV-IO$^+$ to their provenance solutions. We are in communication with the researchers to extend the real-world impact of PROV-IO$^+$ further.



{Similarly, while the current prototype supports 
POSIX and HDF5 I/O  transparently and is extensible by design, there are other popular I/O systems in HPC  (e.g., ADIOS~\cite{ADIOS}) which we have not integrated yet. 
We leave the integration with other I/O libraries as future work.}

{In addition, there are other important aspects of provenance (e.g., security~\cite{TrustworthyUSENIXSec15}) which cannot be ignored in practice. 
We hope that our efforts and the resulting open-source tool  
can facilitate follow-up  research in the communities 
and help address the grand challenge of provenance support for scientific data in general.}

\section{Related Work}
\label{sec:related}

\smallskip
\noindent 
{\bf Database Provenance}. Historically, provenance has been well studied  in  databases  to understand the causal relationship between  materialized views and  table updates~\cite{CuiY-TracingTheLinegeofViewData-TODS2000,whyandwhere}. The concept has also been extended to other usages~\cite{WindomJ-Trio-2004,ProvenanceandProbabilities}. 
In general, database provenance may leverage the well-defined relational model and the relatively strict transformations to capture precise provenance within the system~\cite{Margo-PrimerOnProvenance-ACMQueue14}, which is not applicable for general software. On the other hand, some query optimizations  (e.g., provenance reduction~\cite{DanielDeutch-HypotheticalReasoningSIGMOD19}) could potentially be applied to PROV-IO$^+$. Therefore, 
PROV-IO$^+$ and these tools are  complementary.

\smallskip
\noindent 
{\bf OS-Level Provenance}. Great  efforts have also been made to capture provenance at the operating system (OS) level~\cite{PASS,Margo-Layering-ATC09,SPADE-ProvenanceAuditing-Middleware12}. For example, PASS~\cite{PASS,Margo-Layering-ATC09} intercepts system calls via custom kernel modules for inferring data dependencies. 
Similarly to these efforts, PROV-IO$^+$ recognizes the importance of I/O  syscalls. 
But different from PASS, PROV-IO$^+$ is non-intrusive to the OS kernel. Moreover, PROV-IO$^+$ leverages the unique characteristics of HPC workflows and systems to meet the needs of domain scientists, while PASS is largely inapplicable in this context. More specifically, we elaborate on five key differences as follows:  

(1) \textit{Provenance Model}: PROV-IO$^+$ follows the W3C  specifications to  represent rich provenance information in a relational model (\S\ref{sec:model}). 
In contrast, 
PASS follows the conventional logging mechanism without a general relational model, which limits its capability of capturing and describing complex provenance. For example, PASS has to establish the dependencies among events via a kernel-level logger (i.e., `Observer' ~\cite{Margo-Layering-ATC09}) which cannot interpret the semantics or relations of HPC I/O library events.
Consequently, PASS can only answer relatively limited queries (e.g., ancestor of a node~\cite{Margo-Layering-ATC09}) instead of the rich lineage defined in W3C. 

(2) \textit{System Architecture}:  PROV-IO$^+$ is a user-level solution designed for the HPC environment (\S\ref{sec:architecture}). In contrast, PASS heavily relies on customized kernel modules to achieve its core functionalities. This kernel-based architecture makes PASS incompatible with modern HPC systems. For example,  neither the PASTA file system (in PASS~\cite{PASS}) nor the Lasagna file system (in PASSv2~\cite{Margo-Layering-ATC09}) is compatible with the Lustre PFS  dominant in HPC. In other words, translating the core functionalities of PASS to HPC systems would require substantial efforts (if possible at all), and the implications on performance and scalability is unclear.

(3) \textit{Granularity}: PROV-IO$^+$ can handle fine-grained I/O provenance which is critical for understanding HPC workflows (e.g., the lineage of an attribute of an HDF5 file), while PASS  collects relatively coarse-grained events (e.g., access to an entire  file).

(4) \textit{Tracking APIs:} By embedding in popular HPC I/O libraries, PROV-IO$^+$ does not require modifying the source code to track  I/O provenance.
In contrast, to use PASS, users must consider how to apply six low-level calls (e.g., \texttt{pass\_read},  \texttt{pass\_mkobj}~\cite{Margo-Layering-ATC09}) to the target applications.

(5) \textit{Storage \& Query}: Based on the well-defined model, PROV-IO$^+$ stores provenance as RDF triples backed by the parallel file system. In contrast, PASS relies on its own local file system to generate provenance as local logs.
The storage representation directly affects the user query capability. For example, PROV-IO$^+$  supports querying RDF triples via  SPARQL~\cite{SPARQL}, while PASS only supports a special Path Query Language which is much less popular today. 

 In summary, while PROV-IO$^+$ is partially inspired by the seminal PASS  designed more than a decade ago, the two works are  different due to the different goals and contexts. Therefore, we view PASS and PROV-IO$^+$ as complementary.


\smallskip
\noindent 
{\bf Workflow \& Application Provenance}. Provenance models or systems for workflows and/or applications have also been explored~\cite{Karma, PROV-ML,ProvLake, MiddlewareOnlyProvenance}. 
For example, Karma~\cite{Karma} describes a model with a hierarchy of ‘workflow-service-application-data’. However, the model is designed for the cloud environment and cannot cover diverse HPC needs (e.g., HDF5 attributes, MPI ranks).
PROV-ML~\cite{PROV-ML} is a series of well-defined specifications for machine learning workflows. 
Different from PROV-ML, PROV-IO$^+$ is designed for general HPC workflows.
IBM ProvLake~\cite{ProvLake} is a lineage data management system capable of capturing data provenance across programs. 
Unlike PROV-IO$^+$,
ProvLake always require users to modify the source code  using its special APIs, which severely limits its usage and scalability for complicated HPC workflows.
Similar to PROV-IO$^+$, there are a few provenance capturing tools using DBMS to store queriable provenance data,  but they do not follow any widely used provenance models~\cite{Braid-DB, Chimbuko, Horde-ICDCS}. 

\smallskip
\noindent 
{\bf Other Usage of Provenance}.
Provenance has been applied to other venues. 
For example, MOLLY uses lineage-driven fault injection to expose bugs in fault-tolerant protocols 
~\cite{PeterAlvaro-LineargeDriveFaultInjection-SIGMOD15}.
There have been a multitude of domain-specific or application-specific provenance and ontology management implementations. However, they do not capture the I/O access information that PROV-IO$^+$ manages. 
We believe the comprehensive provenance information enabled by PROV-IO$^+$ 
can also be leveraged to stimulate several data quality and storage optimizations, 
which we leave as future work.

\smallskip
\noindent 
{\bf Non-Provenance Tools}. In addition, great efforts have been made to manage workflows~\cite{Taverna, Effis-WDMapp} or log I/O events for various purposes~\cite{Darshan, Recorder-pdsw,zheng2014torturing,cao2016generic,cao2018pfault,han2021study,gatla2017understanding,gatla2018towards-fast18,gatla2018towards-tos,zhang2021sentilog,dai2019performance,xu2018understanding,xu2019lessons,shi2017command,ZhengGMProf,huang2013liu,zhang2021benchmarking}. While they are effective for their original goals, they are insufficient to address  provenance needs in general due to a number of reasons: 
(1) no relational model to support tracking  or querying rich provenance (e.g.,  various relations defined in W3C PROV-DM~\cite{TheW3CPROVfamily}); 
(2) agnostic to the fine-grained semantics in  HPC I/O libraries (e.g., HDF5 attributes);
(3) little portability across different I/O libraries or workflow environments; 
(4) no programmable interface to specify customized provenance needs.




\section{Conclusion \& Future Work}
\label{sec:conclude}
We have introduced a  provenance tool called PROV-IO$^+$  for scientific data on HPC systems.  Experiments with representative HPC workflows show that PROV-IO$^+$ can address diverse provenance needs  with reasonable overhead. We believe that PROV-IO$^+$ represents a promising direction toward ensuring the rigorousness and trustworthiness of scientific data management.
In the future, we will address the limitations mentioned in \S\ref{sec:discussion}. 
Moreover, in the Top Reco case studied in this paper, the domain scientists would like to identify the best configurations across multiple runs of the workflow.
In other words, there is a need of provenance across multiple executions of the same workflow. 
Similar cross-workflow provenance may be needed when multiple different workflows cooperate to process shared datasets, 
which requires additional modeling and interface to bridge the semantic gap between workflows.  We would like to investigate such complex multi-workflow scenario as well.
In addition, we observe that diagnosing the correctness and performance anomalies in HPC systems is increasingly challenging due to the complexity (e.g., a single SSD  failure may cause the ``blast radius'' problem due to system dependencies), and we will apply PROV-IO$^+$  to address such open challenges in the future.
Overall, we believe that PROV-IO$^+$ represents a promising direction toward ensuring the rigorousness and trustworthiness of scientific data management.



    \vspace{-0.1in}
\section{Acknowledgments}
\label{sec:acknowledgments}

The authors would like to thank 
the
anonymous reviewers for the constructive feedback.  
We also thank Xiangyang Ju for providing the Top Reco workflow, and James Loo, Jim Wayda for the help on Samsung systems. This work was supported in part by NSF under grants CNS-1855565, CCF-1853714, CCF-1910747  and CNS-1943204. Any opinions, findings, and conclusions expressed in this material are those of the authors and do not necessarily reflect the views of the sponsors. 
This manuscript has been authored by an author at Lawrence Berkeley National
Laboratory under Contract No. DE-AC02-05CH11231 with the U.S. Department of
Energy. The U.S. Government retains, and the publisher, by accepting the article for
publication, acknowledges, that the U.S. Government retains a non-exclusive, paid-up,
irrevocable, world-wide license to publish or reproduce the published form of this
manuscript, or allow others to do so, for U.S. Government purposes. 

% ...


\printbibliography

\end{document}